\newif\iflocal
\localfalse

\newif\ifuseapj
\useapjtrue

\pdfoutput=1

\ifuseapj
\documentclass[iop, apj, twocolappendix, numberedappendix]{emulateapj}
\else
\documentclass[fleqn, usenatbib, usedcolumn]{mnras}
\fi

\usepackage{graphicx}
\usepackage{xspace}
\usepackage{amsmath}
\usepackage{framed}
\usepackage{txfonts}
\usepackage{epstopdf}
\usepackage{color}
\usepackage{rotating}
\usepackage{natbib}
\usepackage{ulem}
\usepackage{xspace}
\usepackage[colorlinks=true,urlcolor=blue,linkcolor=blue,citecolor=blue]{hyperref}
\usepackage{sistyle}

\iflocal
\def\includedir{/Users/diemer/University/docs/latex}
\def\figdir{figs}
\else
\def\includedir{.}
\def\figdir{figs}
\fi

\def\gtsima{$\; \buildrel > \over \sim \;$}
\def\ltsima{$\; \buildrel < \over \sim \;$}
\def\prosima{$\; \buildrel \propto \over \sim \;$}
\def\gsim{\lower.7ex\hbox{\gtsima}}
\def\lsim{\lower.7ex\hbox{\ltsima}}
\def\simgt{\lower.7ex\hbox{\gtsima}}
\def\simlt{\lower.7ex\hbox{\ltsima}}
\def\simpr{\lower.7ex\hbox{\prosima}}

\newcommand{\dpar}[2]{\frac{\partial #1}{\partial #2}}
\newcommand{\dparinl}[2]{\partial #1 / \partial #2}

\newcommand{\vect}[1]{{\pmb #1}}

\newcommand{\dotp}{\boldsymbol{\cdot}}
\newcommand{\grad}{\boldsymbol{\nabla}}
\newcommand{\lapl}{\grad^2}
\newcommand{\divg}{\grad \dotp}

\newcommand{\km}{{\rm km}}

\newcommand{\scnd}{{\rm s}}

\newcommand{\mprot}{{m_{\rm p}}}

\newcommand{\boltz}{k_{\rm B}}

\newcommand{\cs}{c_{\rm s}}

\newcommand{\cfl}{\alpha_{\rm cfl}}

\newcommand{\flux}{\mathcal{F}}
\newcommand{\fluxv}{\vect{\flux}}
\newcommand{\vx}{v_{\rm x}}
\newcommand{\vy}{v_{\rm y}}

\def\ulula{\textsc{Ulula}\xspace}

\def\rmr{{\rm r}}

\def\rmC{{\rm C}}

\def\rmL{{\rm L}}

\def\rmR{{\rm R}}

\def\rmX{{\rm X}}

\usepackage{etoolbox}
\makeatletter
\patchcmd{\NAT@citex}
  {\@citea\NAT@hyper@{\NAT@nmfmt{\NAT@nm}\NAT@date}}
  {\@citea\NAT@nmfmt{\NAT@nm}\NAT@hyper@{\NAT@date}}
  {}
  {}
\patchcmd{\NAT@citex}
  {\@citea\NAT@hyper@{%
     \NAT@nmfmt{\NAT@nm}%
     \hyper@natlinkbreak{\NAT@aysep\NAT@spacechar}{\@citeb\@extra@b@citeb}%
     \NAT@date}}
  {\@citea\NAT@nmfmt{\NAT@nm}%
   \NAT@aysep\NAT@spacechar%
   \NAT@hyper@{\NAT@date}}
  {}
  {}
\patchcmd{\NAT@citex}
  {\@citea\NAT@hyper@{%
     \NAT@nmfmt{\NAT@nm}%
     \hyper@natlinkbreak{\NAT@spacechar\NAT@@open\if*#1*\else#1\NAT@spacechar\fi}%
       {\@citeb\@extra@b@citeb}%
     \NAT@date}}
  {\@citea\NAT@nmfmt{\NAT@nm}%
   \NAT@spacechar\NAT@@open\if*#1*\else#1\NAT@spacechar\fi%
   \NAT@hyper@{\NAT@date}}
  {}
  {}
\makeatother

\ifuseapj

\shorttitle{ULULA}
\shortauthors{Diemer}

\slugcomment{ }

\else

\title[ULULA]{ULULA: An ultra-lightweight 2D hydrodynamics code for teaching and experimentation}

\author[Diemer]{Benedikt Diemer\thanks{Email: \href{mailto:diemer@umd.edu}{diemer@umd.edu}}
\vspace{1mm}
\\
Department of Astronomy, University of Maryland, College Park, MD 20742, USA \\
}

\date{}
\pubyear{2025}

\fi

\begin{document}

\ifuseapj

\title{ULULA: An ultra-lightweight 2D hydrodynamics code for teaching and experimentation}
\author{Benedikt Diemer}
\affil{
Department of Astronomy, University of Maryland, College Park, MD 20742, USA; \href{mailto:diemer@umd.edu}{diemer@umd.edu}
}

\else
\label{firstpage}
\pagerange{\pageref{firstpage}--\pageref{lastpage}}
\maketitle
\fi


\begin{abstract}
Hydrodynamics is a difficult subject to teach in the classroom because most relevant problems must be solved numerically rather than analytically. While there are numerous public hydrodynamics codes, the complexity of production-level software obscures the underlying physics and can be overwhelming to first-time users. Here we present \ulula, an ultra-lightweight python code to solve hydrodynamics and gravity in 2D. The main goal is for the code to be easy to understand, extend, and experiment with. The simulation framework consists of fewer than $800$ active lines of pure python code, but it includes a robust MUSCL-Hancock scheme with exchangeable components such as Riemann solvers, reconstruction schemes, boundary conditions, and equations of state. Numerous well-known hydrodynamics problems are provided and can be run in a few minutes on a laptop. The code is open-source, generously commented, and extensively documented.
\end{abstract}

\ifuseapj


\else

\begin{keywords}
hydrodynamics
\end{keywords}

\fi


\section{Introduction}
\label{sec:intro}

The dominant physical mechanisms in the Universe are arguably gravity and hydrodynamics (HD). The latter largely controls the evolution of diverse systems such as gas planets, stars, the interstellar and intergalactic medium, galaxy clusters, and the primordial Universe. While Newtonian gravity can be taught in high school, HD is a much more technical subject that is often not even included in undergraduate physics curricula. One reason for this absence may be that solving HD problems in the classroom can be difficult and unsatisfying. While the Euler equations are relatively simple in the sense that they do not contain an overwhelming number of symbols, there are few relevant problems with derivable, analytical solutions (such as sound waves or hydrostatic atmospheres). Most modern research problems need to be solved numerically, meaning that pen-and-paper teaching fails to reflect the day-to-day nature of research related to HD systems.

To make matters worse, the learning curve for numerical HD is relatively steep because simple PDE solving techniques fail. For example, finite-differencing schemes can be taught fairly quickly to students with some programming background, but those schemes fail to maintain positivity or conserve mass. The non-linear nature of the Euler equations, as manifested by the ubiquity of discontinuities, has led to a long search for viable solvers that met its first real success with the development of finite-volume schemes \citep{godunov_59}. However, an understanding of how finite-volume solvers work is not easily communicated. The theoretical framework by itself is complex and abstract unless coupled to numerical experimentation, but such practical exploration means either writing a scheme from the ground up (which is too time-consuming for most class environments) or running existing HD codes. Unfortunately, production-level astrophysical HD codes tend to be poor tools for learning and experimentation because the underlying equations are necessarily buried deep in complex code for adaptive grids, parallelization, I/O, and physics other than pure HD. Moreover, compiling C/Fortran code is often challenging enough to make such codes impossible to use in, say, homework sets.

In this work, we present \ulula, an extremely lightweight python HD and gravity code designed to be easily understandable and to present a low barrier to entrance. \ulula was written with a strong focus on the brevity and legibility of the code while providing high-performance, real-life HD algorithms. To this end, the entire HD solver framework fits into a single python file with less than $800$ active lines of code. The limitation to 2D represents a compromise between simplicity and performance: 1D codes quickly get boring because many HD problems are intrinsically multi-dimensional, but 3D problems are too large to be solved in reasonable time without parallelization. For the sake of simplicity, \ulula is written in pure python with \texttt{numpy}, with no extra libraries that need to be installed or compiled (such as \texttt{jax} or \texttt{numba}). Given \ulula's pedagogical focus, the code comes with extensive online documentation and is heavily commented. In no more than two lines of code, the user can run pre-implemented problem setups, experiment with different HD solvers, and create plots and movies.

\ulula is not the first pedagogical HD code written in python. Most notably, \textsc{Pyro} \citep{zingale_14_pyro, harpole_19} was designed with a similar purpose in mind and is paired with an open-source book on numerical HD \citep{zingale_21}. In detail, however, the codes are structured rather differently. While \ulula is built as a single simulation framework with interchangeable solver components, \textsc{Pyro} includes a much larger array of solvers for specific problems, such as advection, multi-grid relaxation, diffusion, and incompressible flows. As a result, the total code volume of \textsc{Pyro} is many times larger than that of \ulula. Moreover, \textsc{Pyro} uses just-in-time calculation to speed up certain routines whereas \ulula is based purely on \texttt{numpy}. 

An even shorter, more easily comprehensible alternative to \ulula and \textsc{Pyro} are short python scripts that address a specific HD-related problem. For example, the ``Create your own'' series of scripts by Philip Mocz (see \href{https://github.com/stars/pmocz/lists/create-your-own}{github repository}) excels in solving complex problems such as the Rayleigh-Taylor instability in as few lines of code as possible. On the other end of the complexity spectrum, the general speedup of python via compiled libraries has enabled the creation of production-level python codes such as \textsc{PySPH} \citep{ramachandran_19} or \textsc{Dedalus} \citep{burns_20}. Unlike \ulula, those codes are used for research and are incomparable in their size and complexity.

The paper is structured as follows. We briefly present the fluid equations in \S\ref{sec:theory}. \S\ref{sec:implementation} gives a detailed description to the algorithms and problem setups that are implemented in \ulula. We test the accuracy and performance of the code in \S\ref{sec:results}. We discuss avenues for experimentation and development in \S\ref{sec:future}. The code is freely available at \href{https://bitbucket.org/bdiemer/ulula}{bitbucket.org/bdiemer/ulula} and documented at \href{https://bdiemer.bitbucket.io/ulula}{bdiemer.bitbucket.io/ulula}.


\section{Theory}
\label{sec:theory}

\ulula solves the Euler equations for compressible, inviscid fluids under gravity, which can be expressed as three conservation laws for mass, momentum, and total energy. The changes in these quantities at a fixed location can be understood as the divergence of a flux plus source terms,
\begin{align}
\label{eq:euler}
\dpar{\rho}{t} &+ \divg (\rho \vect{v}) &=&\ 0 \nonumber \\
\dpar{(\rho \vect{v})}{t} &+ \divg \left( \rho \vect{v} \otimes \vect{v} +  \vect{I} P \right) &=&\ - \rho \grad \Phi\nonumber  \\
\dpar{E}{t} &+ \divg \left([E + P] \vect{v} \right) &=&\ \rho \dpar{\Phi}{t} \,,
\end{align}
where $\rho$ is the density, $\vect{v} = (\vx, \vy)$ is the fluid velocity vector in 2D, $P$ is the pressure, $E = \rho(\vect{v}^2 / 2 + \varepsilon + \Phi)$ is the total energy per unit volume, $\varepsilon$ is the internal energy per unit mass, and $\Phi$ is the gravitational potential. We make the notation more compact by combining the three equations into a single vector equation for a conservation law,
\begin{equation}
\label{eq:euler_vector}
\dpar{\vect{U}}{t} + \divg \fluxv(\vect{U}) = \vect{S} \,,
\end{equation}
where we have defined the vector of conserved fluid quantities, $\vect{U}$, the vector of flux terms, $\fluxv(\vect{U})$, and the vector of source terms, $\vect{S}$. In addition, we define a vector of ``primitive'' variables $\vect{V}$. For a 2D system, we have 
\begin{equation}
\label{eq:vectors}
\vect{V} \equiv
\left(\begin{array}{c}
\rho \\ \noalign{\medskip}
\vx \\ \noalign{\medskip}
\vy \\ \noalign{\medskip}
P
\end{array}\right) \nonumber 
\qquad
\vect{U} \equiv
\left(\begin{array}{c}
\rho \\ \noalign{\medskip}
\rho \vx \\ \noalign{\medskip}
\rho \vy \\ \noalign{\medskip}
E
\end{array}\right) \nonumber 
\qquad
\vect{S} \equiv
\left(\begin{array}{c}
0 \\ \noalign{\medskip}
- \rho\ \dparinl{\Phi}{x} \\ \noalign{\medskip}
- \rho\ \dparinl{\Phi}{y} \\ \noalign{\medskip}
\rho\ \dparinl{\Phi}{t}
\end{array}\right) 
\end{equation}
and, writing out the velocity flux components in the flux vector, 
\begin{equation}
\label{eq:flux_vector}
\fluxv(\vect{U})  \equiv
\left(\begin{array}{c}
\rho \vect{v} \\ \noalign{\medskip}
(\rho \vx^2 + P,\ \rho \vx \vy) \\ \noalign{\medskip}
(\rho \vx \vy,\ \rho \vy^2 + P) \\ \noalign{\medskip}
(E + P) \vect{v}
\end{array}\right) \,.
\end{equation}
The system of equations must be closed using an equation of state that relates pressure and internal energy. For an ideal gas, we have 
\begin{equation}
P = n \boltz T = \frac{\rho \boltz T}{\mu \mprot} = \rho \varepsilon (\gamma - 1) \,,
\end{equation}
where $n$ is the number density, $\boltz$ is the Boltzmann constant, $T$ is the temperature in Kelvin, $\mu$ is the mean particle mass in proton masses, $\mprot$ is the proton mass, and $\gamma$ is the adiabatic index of the gas. The sound speed in an ideal gas is $\cs = \sqrt{\gamma P / \rho}$. In addition to the ideal gas equation of state, \ulula also implements an isothermal equation of state, where $T$ (and thus $\varepsilon$) are set to a constant value. This case is an example of a ``barotropic'' equation of state, where pressure depends only on density. For such systems, the conservation law for total energy in eq.~\ref{eq:euler} can be omitted since the total energy directly follows from density and internal energy. The sound speed for an isothermal gas is $\cs = \sqrt{P / \rho}$.

Finally, the gravitational potential $\Phi$ can either be imposed as an input variable or be the result of the self-gravity of the fluid, in which case it is governed by the Poisson equation,
\begin{equation}
\lapl \Phi = 4 \pi G \rho \,,
\end{equation}
where $G$ is the gravitational constant. In addition to self-gravity, \ulula allows for user-defined, fixed potentials to capture constant accelerations or potentials due to a mass distribution that is not being simulated hydrodynamically.


\section{Implementation}
\label{sec:implementation}

The \ulula hydro solver is fundamentally a \citet{godunov_59} scheme, meaning that it solves eq.~\ref{eq:euler_vector} by shifting the conserved quantities (namely mass, momentum, and energy) across cells (\S\ref{sec:implementation:timestepping}). The fluxes of these quantities across cell interfaces are calculated by means of a Riemann solver (\S\ref{sec:implementation:riemann}). The fluid states that serve as input to the Riemann solver can be interpolated in space (\S\ref{sec:implementation:spatial}) and/or time (\S\ref{sec:implementation:time}) to increase the accuracy of the scheme. We discuss our implementation of gravity in \S\ref{sec:implementation:gravity}. We summarize the overall architecture of the code in \S\ref{sec:implementation:code}.

The algorithms described here are not new. Similar solvers have been described in numerous papers and textbooks \citep[e.g.,][]{toro_09, zingale_21}. The following description aims to be pedagogical and to closely follow the expressions used in \ulula. The choices for various numerical components are summarized in Table~\ref{table:sims}.

\begin{table}
\centering
\caption{Implementation choices of components in the \ulula hydro solver.}
\label{table:sims}
\begin{tabular}{llc}
\hline
Category/Name & Comments & \S/eq. \\
\hline
{\bf Riemann solvers} &  & \ref{sec:implementation:riemann}  \\
HLL & Ignores contact discontinuities & \ref{eq:hll} \\
HLLC & HLL plus contact discontinuities & \ref{eq:hllc} \\
{\bf Spatial interpolation} & & \ref{sec:implementation:spatial} \\
None (constant) & 1st-order & \ref{eq:reconstruction:constant} \\
Linear + MinMod & 2nd-order, conservative limiter & \ref{eq:limiter:minmod} \\
Linear + van Leer & 2nd-order, intermediate limiter & \ref{eq:limiter:vanleer} \\
Linear + MC & 2nd-order, aggressive limiter & \ref{eq:limiter:mc} \\
{\bf Time interpolation} & & \ref{sec:implementation:time} \\
None (Euler integrator) & 1st-order and unstable &  \ref{eq:euler_integration} \\
Hancock (conserved) & 2nd-order, edge states at $t = n+1/2$ & \ref{eq:hancock_cons} \\
Hancock (primitive) & 2nd-order, faster than cons. Hancock & \ref{eq:hancock_prim} \\
\hline
\end{tabular}
\end{table}

\subsection{Timestepping}
\label{sec:implementation:timestepping}

We denote cells by their index $i$ in the $x$-direction and index $j$ in the $y$-direction. In the most general terms, a Godunov scheme updates a cell over a time interval $\Delta t$ from timestep $n$ to timestep $n + 1$ by adding to and subtracting from the conserved quantities. In two dimensions, this operation reads
\begin{align}
\label{eq:godunov}
\vect{U}_{ij}^{n+1} = \vect{U}_{ij}^n &+ \frac{\Delta t}{\Delta x} \left( \fluxv_{i-1/2}^{n+1/2} - \fluxv_{i+1/2}^{n+1/2} \right) \nonumber \\
&+ \frac{\Delta t}{\Delta x} \left( \fluxv_{j-1/2}^{n+1/2} - \fluxv_{j+1/2}^{n+1/2} \right) + \Delta t\ \vect{S_{ij}^{n+1/2} } \,,
\end{align}
where $\Delta x$ is the spatial extent of the cells (which all share the same size in \ulula). The flux at $i - 1/2$ refers to the interface between cells $i$ and $i - 1$, the flux at $i + 1/2$ refers to the interface between cells $i$ and $i + 1$, and so on for $j$. The differencing of the flux vectors in eq.~\ref{eq:godunov} corresponds to a discretized version of the divergence in eq.~\ref{eq:euler_vector}. We have marked the flux and source terms to be taken at time $n + 1/2$ to highlight that they should, ideally, approximate their integrated values over a timestep $\Delta t$. An accurate scheme boils down to finding accurate expressions for these time-averaged fluxes at the cell interfaces, as well as for the time-averaged source terms. Specifically, we will choose schemes to be 2nd-order accurate, meaning that all error terms that are proportional to $\Delta x$ or $\Delta t$ should cancel. However, \ulula does also implement first-order schemes for experimentation.

As written, eq.~\ref{eq:godunov} implies that we need to evaluate the flux terms in both dimensions based on the current fluid state $\vect{U}^{n}$ before adding and subtracting them. In constructing the fluxes at the interface, we also need to worry about the transverse fluxes, for example the $y$-component of the flux vectors $\fluxv_{i-1/2}^{n+1/2}$ (eq.~\ref{eq:flux_vector}). Such a scheme is known as an ``unsplit'' solver, which has the advantage that the spatial dimensions are treated symmetrically. 

In \ulula, however, we choose a simpler ``split'' scheme, where we compute and apply the fluxes in the $x$ and $y$ directions separately. During each of these ``sweeps,'' we care only about the flux in the given sweep direction while ignoring transverse fluxes. The fluid state is then updated by the flux differences in eq.~\ref{eq:godunov} before the second sweep is performed. Mathematically speaking, we can think of this process as splitting the full flux vector of eq.~\ref{eq:flux_vector} into two components,
\begin{equation}
\label{eq:flux_vector_split}
\fluxv_{\rm x}(\vect{U}_{ij})  \equiv
\left(\begin{array}{c}
\rho \vx \\ \noalign{\medskip}
\rho \vx^2 + P \\ \noalign{\medskip}
\rho \vx \vy \\ \noalign{\medskip}
(E + P) \vx
\end{array}\right) 
\qquad
\fluxv_{\rm y}(\vect{U}_{ij})  \equiv
\left(\begin{array}{c}
\rho \vy \\ \noalign{\medskip}
\rho \vy \vx \\ \noalign{\medskip}
\rho \vy^2 + P \\ \noalign{\medskip}
(E + P) \vy
\end{array}\right) 
\end{equation}
where the subscripts $ij$ remind us that these are flux vectors for the fluid state in a given cell as opposed to the fluxes across cell interfaces that we need to compute for eq.~\ref{eq:godunov}. None the less, we have hugely simplified the problem because we can write code to solve a one-dimensional problem and execute it twice in the different dimensions. In practice, we apply the same routines to $\fluxv_{\rm x}$ and $\fluxv_{\rm y}$ but switch $\vx$ and $\vy$ as well as the second and third indices. It turns out that split schemes maintain second-order accuracy as long as the sweeps are executed in an order such as $x$-$y$-$y$-$x$-$x$-$y$ and so on \citep{strang_68}, which is adopted in \ulula.

One complication that arises in split solvers is that the fluid state changes between sweeps, which can change the size of the timestep. The timestep is limited by the \citet[][hereafter CFL]{courant_28} condition, which ensures causality in the sense that information cannot travel by more than one cell per timestep. To be conservative, we set the timestep to
\begin{equation}
\Delta t = \cfl \frac{\Delta x}{c_{\rm max}} \,,
\end{equation}
where $0 \leq \cfl \leq 1$ is the CFL number and 
\begin{equation}
c_{\rm max} \equiv {\rm max}( |\vx| + \cs, |\vy| + \cs)
\end{equation}
is the maximum speed anywhere in the simulation volume (which can be derived as the eigenvalues of matrices such as the one in eq.~\ref{eq:jacobian}). This maximum speed may well increase after the first sweep, but we must maintain the same $\Delta t$ in both sweeps for the cancellation of first-order error terms to work. We thus allow second sweeps with timesteps corresponding to a maximum CFL number $\alpha_{\rm max}$ that is larger than $\cfl$. If the timestep would correspond to a CFL number larger than $\alpha_{\rm max}$, we reset to the state before the first sweep and try again with a reduced timestep. This process is repeated until we successfully complete both sweeps.

Finally, we need to approximate the time integral over the source term in eq.~\ref{eq:godunov}. We achieve second-order accuracy by splitting it in two half-steps,
\begin{equation}
\Delta t\ \vect{S_{ij}^{n+1/2} } \approx \frac{\Delta t}{2} \vect{S}_{ij}^{n} + \frac{\Delta t}{2} \vect{S}_{ij}^{n+1} \,.
\end{equation}
In practice, this approximation is computed by applying the source term corresponding to the current fluid state before and after the sweeps, respectively.

\subsection{Riemann solver}
\label{sec:implementation:riemann}

\begin{figure*}
\centering
\includegraphics[trim =  18mm 18mm 3mm 9mm, clip, width=0.24\textwidth]{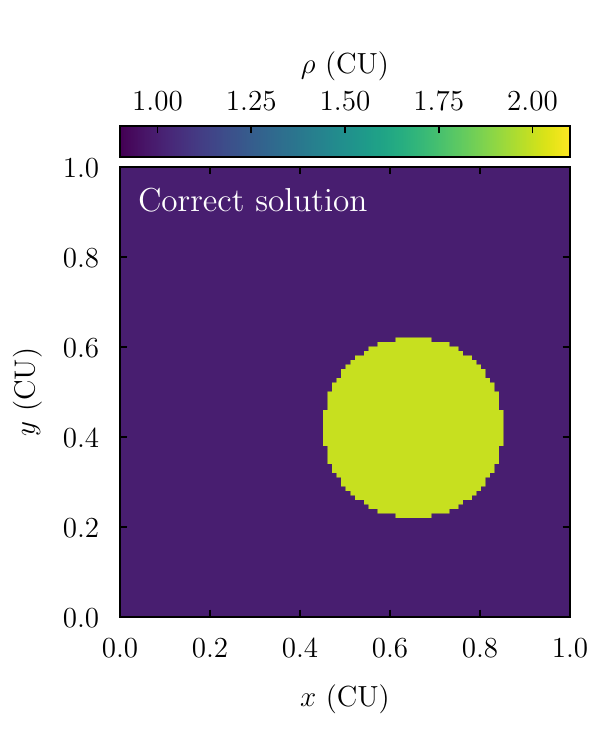}
\includegraphics[trim =  18mm 18mm 3mm 9mm, clip, width=0.24\textwidth]{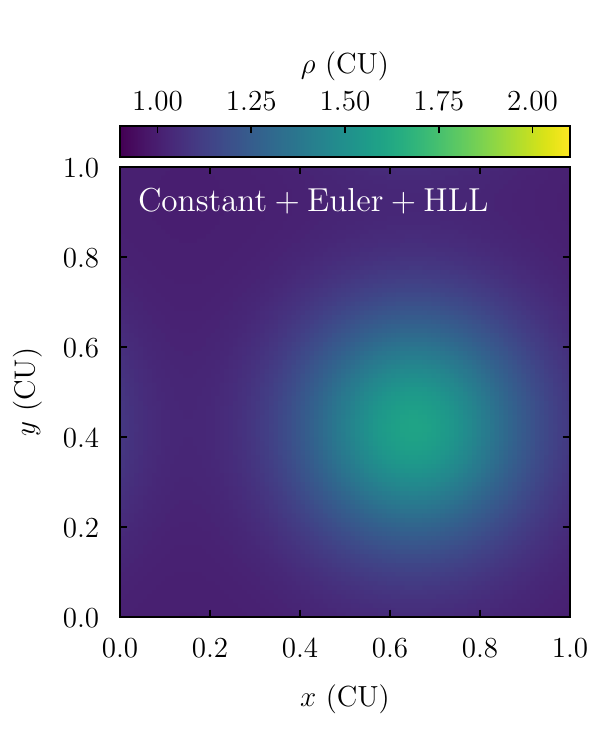}
\includegraphics[trim =  18mm 18mm 3mm 9mm, clip, width=0.24\textwidth]{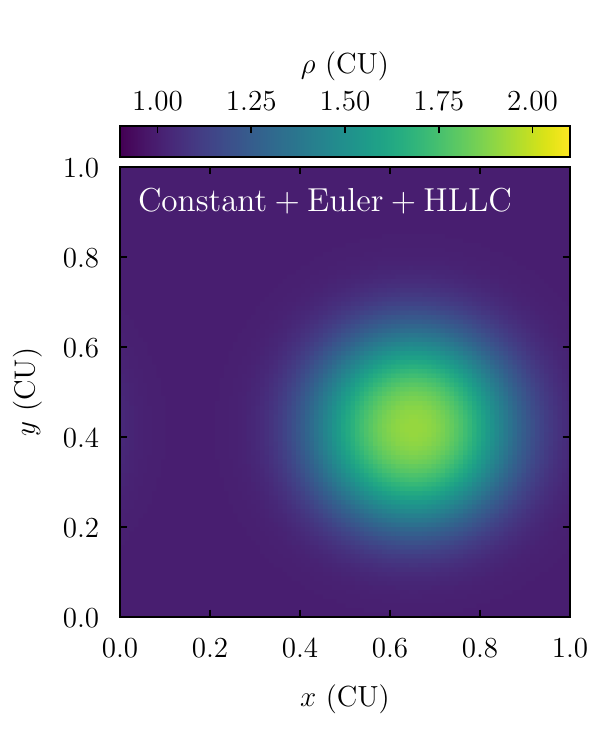}
\includegraphics[trim =  18mm 18mm 3mm 9mm, clip, width=0.24\textwidth]{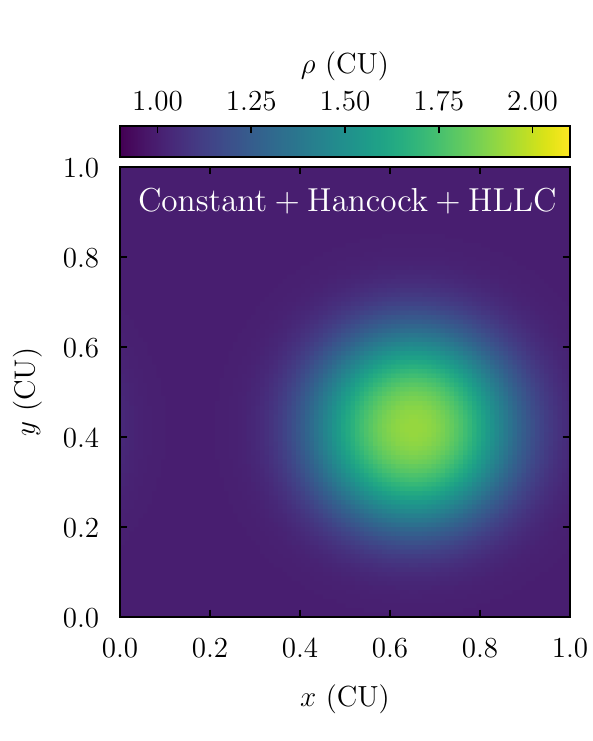}
\includegraphics[trim =  18mm 20mm 3mm 28mm, clip, width=0.24\textwidth]{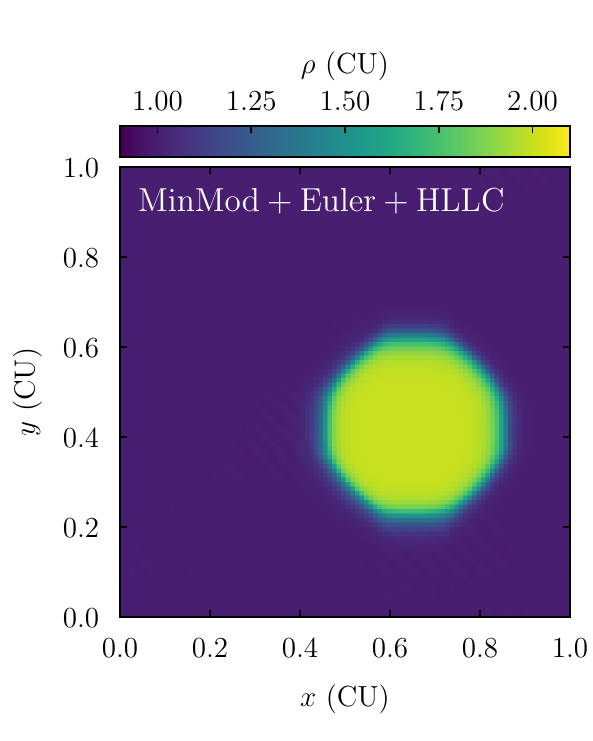}
\includegraphics[trim =  18mm 20mm 3mm 28mm, clip, width=0.24\textwidth]{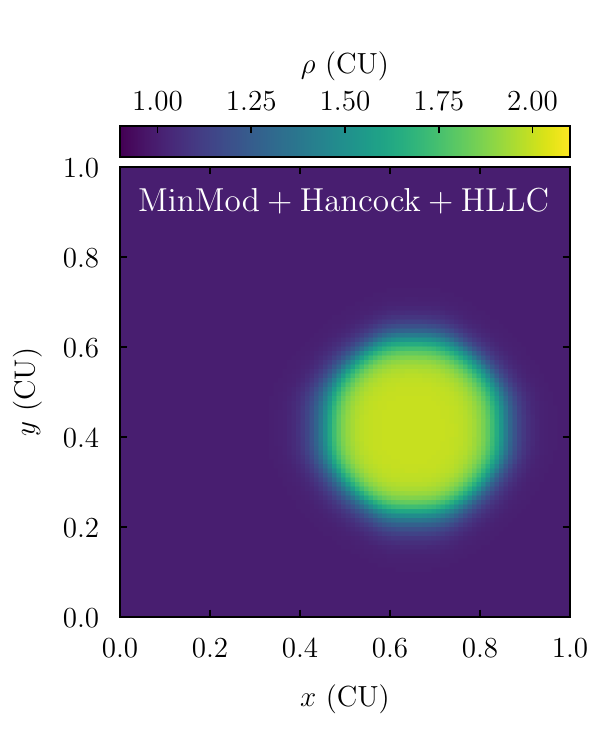}
\includegraphics[trim =  18mm 20mm 3mm 28mm, clip, width=0.24\textwidth]{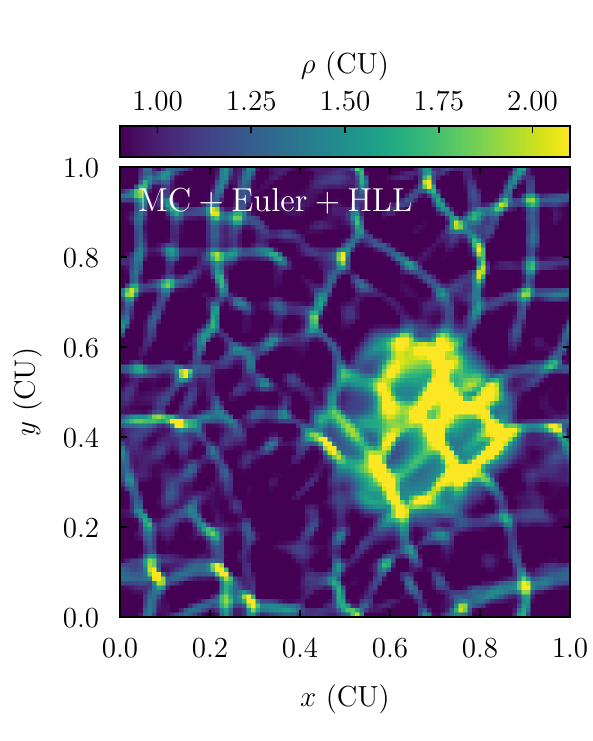}
\includegraphics[trim =  18mm 20mm 3mm 28mm, clip, width=0.24\textwidth]{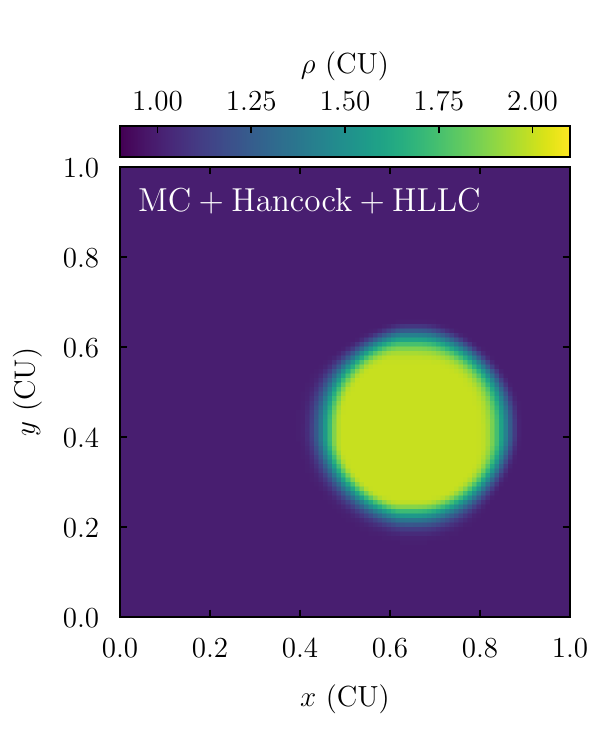}
\caption{Advection test of HD algorithms. A sharply delineated blob of dense gas is initially placed at the center of the domain and transported across the domain with speed $\vx = 0.5$ and $\vy = 0.4$. We analyze the results when the blob has roughly crossed the domain once, after about $550$ time steps. An ideal scheme would perfectly preserve the blob (top left panel). The simplest scheme, a first-order ``Euler'' time integration scheme with no spatial interpolation and the HLL Riemann solver (top second panel), is extremely diffusive. Using the HLLC Riemann solver helps significantly in this particular test because of the contact discontinuity at the edge of the blob (top third panel). In this case, using Hancock time-interpolated states does not make a huge difference (top right panel). Adding spatial interpolation and the conservative MinMod slope limiter improves the solution significantly, although the blob shows deformations when using the first-order Euler timestepping scheme (bottom left panel). Using Hancock timestepping is more diffusive but also reduces those artifacts (bottom second panel). When combining the more aggressive MC slope with Euler timestepping leads to instability (bottom third panel). However, the Hancock time interpolation restores stability and leads to the best solution (bottom right panel).}
\label{fig:advection}
\end{figure*}

It is not immediately obvious how to compute the fluxes across cell interfaces that appear in eq.~\ref{eq:godunov} from the cell states $\vect{U}_{ij}$, or from the corresponding fluxes within the cells, $\fluxv(\vect{U}_{ij})$. One could imagine averaging over the adjacent cells, but this would result in a poor solver because it ignores the physical reality that signals such as shocks travel across interfaces in a particular direction. Such physical effects are taken into account by Riemann solvers, i.e., algorithms that solve for the flux across a boundary between discontinuous states to the left and right, $\vect{U}_\rmL$ and $\vect{U}_\rmR$. Numerous Riemann solvers have been proposed \citep[see][for a thorough treatment]{toro_09}. For the 1D Euler equations, the Riemann problem can be solved numerically to arbitrary accuracy \citep[e.g.,][]{colella_85_cg}, but approximate solutions tend to be computationally faster. In \ulula, we implement the relatively simple HLL Riemann solver (named after \citealt{harten_83}),
\begin{align}
\label{eq:hll}
\fluxv_{i\pm1/2}^{\rm HLL} = 
\begin{cases}
\fluxv_\rmL & \forall \quad S_\rmL \ge 0 \\
\frac{S_\rmR \fluxv_\rmL - S_\rmL \fluxv_\rmR + S_\rmL S_\rmR (\vect{U}_\rmR - \vect{U}_\rmL)}{S_\rmR - S_\rmL} & \mathrm{otherwise}  \\
\fluxv_\rmR & \forall \quad S_\rmR \le 0
\end{cases}
\end{align}
where $\fluxv_\rmL \equiv \fluxv(\vect{U}_\rmL)$, $\fluxv_\rmR \equiv \fluxv(\vect{U}_\rmR)$, $S_\rmL \equiv v_\rmL - \cs$, and $S_\rmR \equiv v_\rmR + \cs$. The physical interpretation is that if even the fastest left-directed wave speed $S_\rmL$ is positive, all waves are traveling to the right and the state at the interface is $\vect{U}_\rmL$ (and vice versa). This is, for example, the case in supersonic flows where $v_\rmL > \cs$. In the intermediate case, we use an averaged flux according to eq.~\ref{eq:hll}. 

This intermediate HLL flux does not distinguish states to the left and right of a contact discontinuity that emerges in the general Riemann problem for the Euler equation. This additional decision is included in more complicated HLLC Riemann solver \citep[HLL + contact; e.g.,][]{toro_94, li_05_hllc}, which we also implement in \ulula. We follow the formulation of \citet{toro_09}, but slightly different versions have also been proposed. The key difference to HLL is that we compute two intermediate states to the left and right of a contact discontinuity that moves with speed $S^\ast$ such that
\begin{align}
\label{eq:hllc}
\fluxv_{i\pm1/2}^{\rm HLLC} = 
\begin{cases}
\fluxv_\rmL & \forall \quad S_\rmL \ge 0 \\
\fluxv_\rmL + S_\rmL (\vect{U}_\rmL^\ast - \vect{U}_\rmL) & \forall \quad  S_\rmL < 0\ {\rm and}\ S^\ast \ge 0 \\
\fluxv_\rmR + S_\rmR (\vect{U}_\rmR^\ast - \vect{U}_\rmR) & \forall \quad S_\rmR > 0\ {\rm and}\ S^\ast \le 0 \\
\fluxv_\rmR & \forall \quad S_\rmR \le 0
\end{cases}
\end{align}
The exact calculation of the wave speeds $S_\rmL$, $S_\rmR$, and $S^\ast$ is fairly complicated \citep{batten_97}. We follow \citet{toro_09} in using the simple estimates of the left-most and right-most waves from the HLL solver, and in estimating the speed of the contact discontinuity as
\begin{equation}
\label{eq:hllc_sstar}
S^\ast = \frac{P_\rmR - P_\rmL + \rho_\rmL v_\rmL (S_\rmL - v_\rmL) - \rho_\rmR v_\rmR (S_\rmR - v_\rmR)}{\rho_\rmL (S_\rmL - v_\rmL) - \rho_\rmR (S_\rmR - v_\rmR)} \,,
\end{equation}
where $v$ represents the velocity in the direction of the sweep (which can be $\vx$ or $\vy$). The density of the intermediate states is
\begin{equation}
\rho_\rmX^\ast = \rho_\rmX \frac{S_\rmX - v_\rmX}{S_\rmX - S^\ast} \,,
\end{equation}
where $X$ is either $L$ or $R$. The full intermediate states are 
\begin{equation}
\label{eq:hllc_ustar}
\vect{U_\rmX^\ast} = 
\left(\begin{array}{c}
\rho_\rmX^\ast \\ \noalign{\medskip}
\rho_\rmX^\ast S^\ast \\ \noalign{\medskip}
\rho_\rmX^\ast v_{\rm X,\perp}\\ \noalign{\medskip}
\frac{\rho_\rmX^\ast E_\rmX}{\rho_\rmX} + \rho_\rmX^\ast (S^\ast - v_\rmX) \left( S^\ast + \frac{P_\rmX}{\rho_\rmX (S_\rmX - v_\rmX)} \right)
\end{array}\right) \,,
\end{equation}
where $v_{\perp}$ refers to the velocity component perpendicular to the sweep, i.e., $\vy$ during an $x$-sweep and $\vx$ during a $y$-sweep. The HLLC solver is computationally more expensive than HLL but also performs demonstrably better in tests that involve contact discontinuities (Fig.~\ref{fig:advection}). Given that pressure and energy explicitly appear in eqs.~\ref{eq:hllc_sstar} and \ref{eq:hllc_ustar}, one would need to reconstruct those quantities when an isothermal EOS is chosen. In this case, HLLC does not offer an improvement over HLL by construction because there are no contact discontinuities in isothermal gas. We thus allow HLLC only for ideal gases.

\subsection{Spatial interpolation}
\label{sec:implementation:spatial}

\begin{figure*}
\centering
\begin{minipage}[l]{0.34\textwidth}
\includegraphics[trim =  0mm 4mm 2mm 0mm, clip, width=\textwidth]{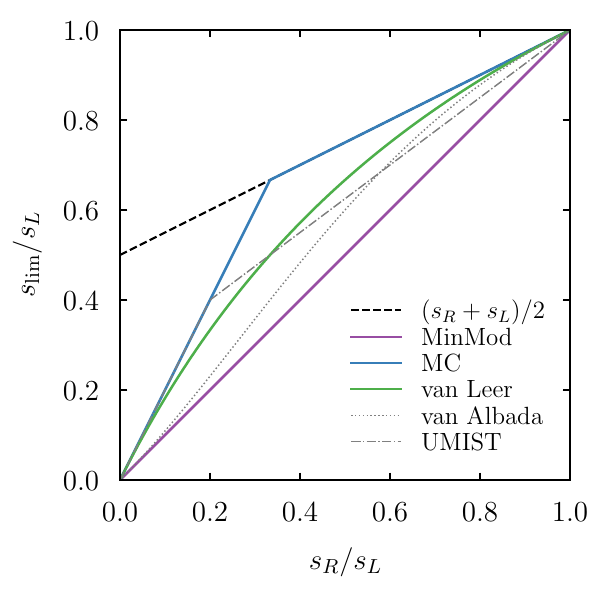}
\end{minipage}
\begin{minipage}[c]{0.65\textwidth}
\includegraphics[trim =  2mm 0mm 2mm 2mm, clip, width=0.49\textwidth]{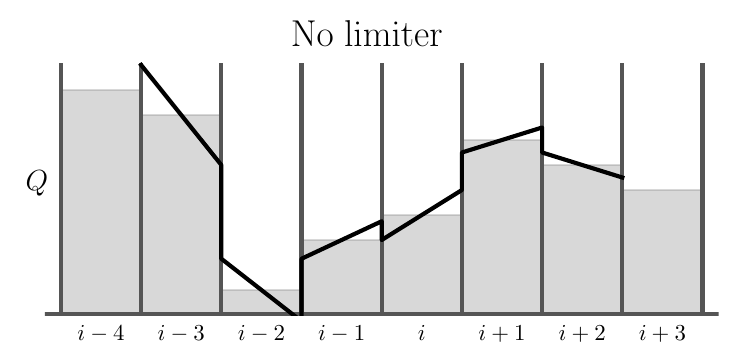}
\includegraphics[trim =  2mm 0mm 2mm 2mm, clip, width=0.49\textwidth]{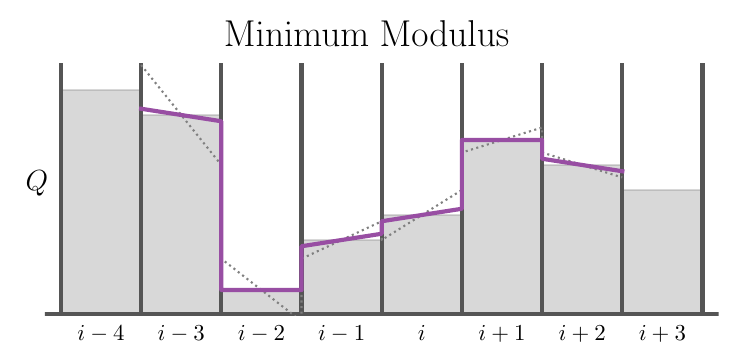}
\includegraphics[trim =  2mm 0mm 2mm 0mm, clip, width=0.49\textwidth]{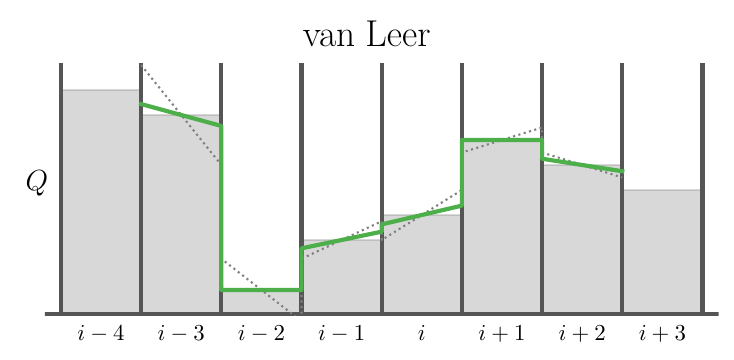}
\includegraphics[trim =  2mm 0mm 2mm 0mm, clip, width=0.49\textwidth]{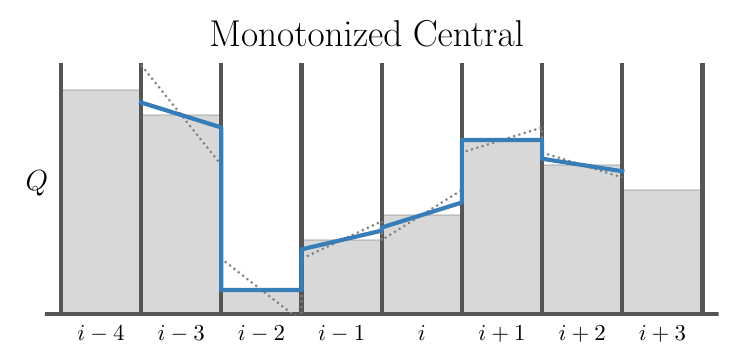}
\end{minipage}
\caption{Visualization of slope limiters and their effect on the reconstruction. The left panel shows the limited slope as  a function of the ratio of the left and right slopes $s_L$ and $s_R$ according to a number of popular limiters. All limiters agree that the limited slope $s_{\rm lim}$ is zero if $s_R$ and $s_L$ have opposite signs or if either of $s_L$ or $s_R$ is zero. All limiters are furthermore symmetric, which allowed us to assume $s_R < s_L$ in the figure without loss of generality. The right panels show the slope limiters applied to an imaginary fluid quantity $Q$. The dotted gray lines show the same solution as the no-limiter panel to highlight the differences. The most conservative limiter is MinMod (purple), which always sets the slope to the smaller of the two absolute values. The most aggressive is MC (monotonized central, blue), which maintains a slope closer to the mean of $s_L$ and $s_R$ as long as they are not too dissimilar.}
\label{fig:limiters}
\end{figure*}

In the previous section, we calculated the fluxes across interfaces from the left and right states $\vect{U}_\rmL$ and $\vect{U}_\rmR$ without specifying how these states should be computed. Let us call $\vect{U}_{\rm i-1/2}$ the state at the left edge of cell $i$ and thus the right side of interface $i - 1/2$, and $\vect{U}_{\rm i+1/2}$ the state at the right edge of cell $i$ and thus the left side of interface $i + 1/2$ (see Fig.~\ref{fig:limiters} for a visualization). The simplest choice is to assume a constant state within each cell, so that 
\begin{equation}
\label{eq:reconstruction:constant}
\vect{U}_{\rm i-1/2} = \vect{U}_{\rm i+1/2} = \vect{U}_{\rm i} \,.
\end{equation}
Such a scheme is manifestly first-order since the error in the solution scales with $\Delta x$. In practice, a constant spatial ``reconstruction'' leads to highly diffusive and unstable schemes (Fig.~\ref{fig:advection}). To achieve second-order accuracy in space, we use a piecewise-linear reconstruction \citep{vanleer_79}. We imagine the fluid variables to vary linearly across each cell with some slope $\vect{s}_{\rm lim}$, where each variable can have a different slope. The state vectors at the left and right edges are then
\begin{equation}
\vect{V}_{\rm i \pm 1/2}^n = \vect{V}_{i}^n \pm \frac{\Delta x}{2}\ \vect{s}_{\rm lim} \,.
\end{equation}
The slope interpolation is generally performed in the primitive variables $\vect{V}$ instead of the conserved variables $\vect{U}$ for reasons that will become apparent shortly. We note that we do not interpolate between cell centers, which would eliminate any differences at the cell edge and obviate the Riemann solver. Instead, we construct the slope from the difference between the neighboring cells. We define the left and right derivatives,
\begin{equation}
\label{eq:reconstruction:slopes}
\vect{s}_\rmL \equiv \frac{\vect{V}_{i} - \vect{V}_{i-1}}{\Delta x} \qquad
\vect{s}_\rmR \equiv \frac{\vect{V}_{i+1} - \vect{V}_{i}}{\Delta x}  \,,
\end{equation}
as well as the central derivative,
\begin{equation}
\vect{s}_\rmC \equiv \frac{\vect{s}_\rmL + \vect{s}_\rmR}{2} = \frac{\vect{V}_{i+1} - \vect{V}_{i-1}}{2 \Delta x} \,.
\end{equation}
However, there is no guarantee that using the central derivative would not lead to unphysical values at the cell edges, such as negative density or pressure (Fig.~\ref{fig:limiters}). We avoid such issues by using a so-called slope limiter, i.e., a function that reduces the central derivative in cases where $\vect{s}_\rmL$ and $\vect{s}_\rmR$ disagree (Fig.~\ref{fig:limiters}). Applying the slope limiter to the primitive variables $\vect{V}$ allows us to directly ensure that there are no negative values or jumps in the pressures.

All limiters fall back to a constant state (zero slope) if the left and right slopes have opposite signs. Moreover, all limiters converge to zero if the smaller of the derivatives is zero. The most conservative limiter, which is called ``minimum modulus'' \citep[MinMod,][]{roe_86}, always chooses the smaller of the two derivatives,
\begin{equation}
\label{eq:limiter:minmod}
s_{\rm MinMod} = {\rm sign}(s_\rmL) \times {\rm min}(|s_\rmL|, |s_\rmR|) \,,
\end{equation}
where we could take the sign of either $s_\rmL$ or $s_\rmR$ since they must, by construction, have the same sign if the limiter is used. The MinMod limiter corresponds to $s_{\rm lim} = s_\rmR$ in the left panel of Fig.~\ref{fig:limiters}, where we have assumed that $s_\rmR \le s_\rmL$. Reasonable limiters can occupy the space between the purple line (MinMod) and the gray dashed line ($s_\rmC$). In \ulula, we implement two more limiters. An intermediate choice is that of \citet{vanleer_74}, which can be expressed as 
\begin{equation}
\label{eq:limiter:vanleer}
s_{\rm vanLeer} = \frac{s_\rmL s_\rmR}{s_\rmC} = 2 \frac{s_\rmL s_\rmR}{s_\rmL + s_\rmR} \,.
\end{equation}
The most aggressive commonly used limiter is called ``monotonized central'' \citep[MC,][]{vanleer_77}. Here we choose the central slope as long as the smaller slope is at least $1/3$ of the larger slope, at which point it changes to twice the smaller slope. Mathematically, this can be expressed as
\begin{equation}
\label{eq:limiter:mc}
s_{\rm MC} = {\rm sign}(s_\rmL) \times {\rm min} \left( 2\ {\rm min}(|s_\rmL|, |s_\rmR|), |s_\rmC| \right) \,,
\end{equation}
which results in the blue line in Fig.~\ref{fig:limiters}. There are many other limiters that fall into the space between MinMod and MC, such as ``van Albada'' \citep{vanalbada_82} or ``UMIST'' \citep{lien_94}. These limiters are shown as thin gray lines in Fig.~\ref{fig:limiters}, but they are not implemented in \ulula because they give results similar to the intermediate van Leer limiter.

While the details of slope limiting may sound somewhat technical, Fig.~\ref{fig:advection} demonstrates that it can have a large impact. For example, a first-order timestepping scheme (\S\ref{sec:implementation:time}) diffuses the advected blob with no reconstruction, yields a sharp but slightly deformed blob with MinMod, and becomes unstable with MC. Similar differences are visible when using a 2nd-order time integration scheme (also discussed in \S\ref{sec:implementation:time}), with MC giving the sharpest results. The choice of limiter depends on the problem though. For example, MC can lead to edge effects in a shock tube test.

\subsection{Time interpolation}
\label{sec:implementation:time}

\begin{figure*}
\centering
\includegraphics[trim =  18mm 20mm 3mm 8mm, clip, width=0.33\textwidth]{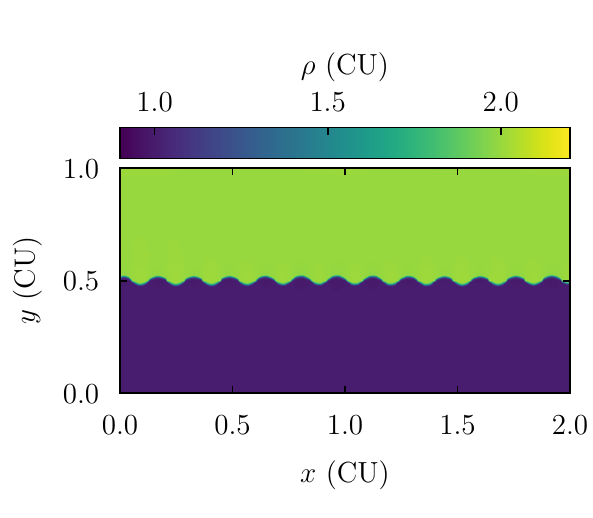}
\includegraphics[trim =  18mm 20mm 3mm 8mm, clip, width=0.33\textwidth]{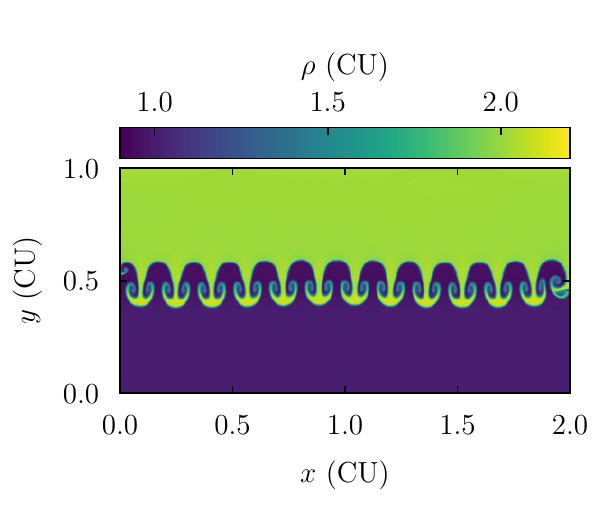}
\includegraphics[trim =  18mm 20mm 3mm 8mm, clip, width=0.33\textwidth]{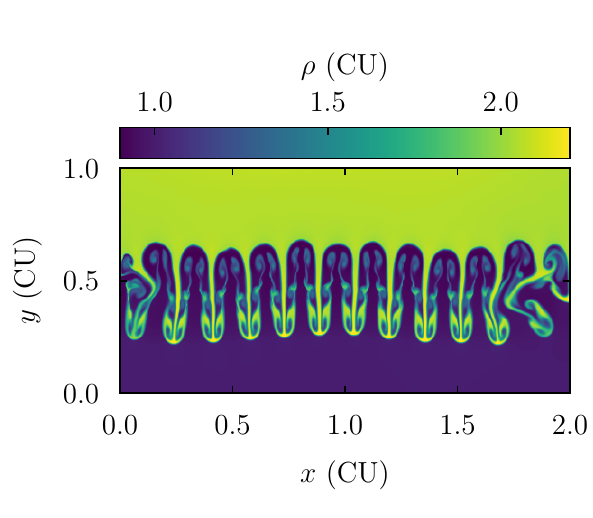}
\includegraphics[trim =  18mm 20mm 3mm 28mm, clip, width=0.33\textwidth]{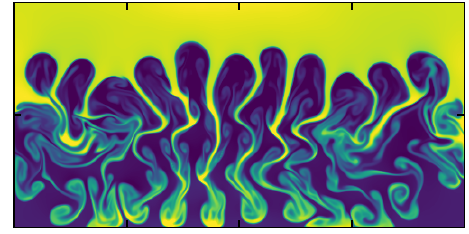}
\includegraphics[trim =  18mm 20mm 3mm 28mm, clip, width=0.33\textwidth]{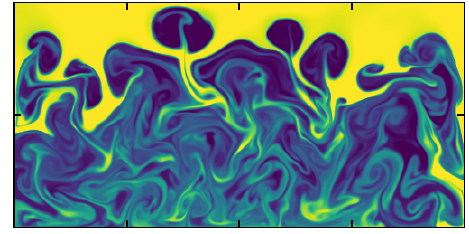}
\includegraphics[trim =  18mm 20mm 3mm 28mm, clip, width=0.33\textwidth]{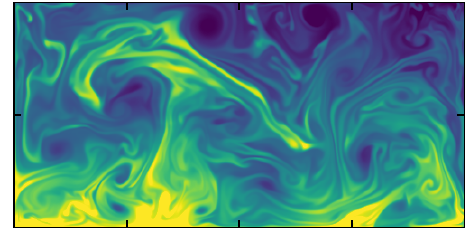}
\caption{Rayleigh-Taylor instability at times $t = 1$, $2$, $3$, $4$, $5$, and $10$ in code units. The initially sharp delineation between a higher density on top and a lower density on the bottom is perturbed by a small $y$-velocity at the center of the domain. The velocity amplitude follows two super-imposed sine waves of equal amplitude, one with a frequency of about $1.5$ wavelengths across the box and one with about $12$. This setup demonstrates that, at fixed amplitude, short wave modes are more Rayleigh-Taylor unstable than long wave modes (leading to only the short mode being visible in the top left panel). However, the long mode eventually does go unstable too and interacts with the short mode (and boundary conditions) to create a complex pattern.}
\label{fig:rt}
\end{figure*}

While our linear reconstruction schemes are now 2nd-order in space, we have yet to achieve the same in time, where the fluxes across interfaces in eq.~\ref{eq:godunov} are meant to represent an average over the timestep. The simplest imaginable scheme is to calculated fluxes based on the current fluid state, meaning that the input states to the Riemann solver are
\begin{equation}
\label{eq:euler_integration}
\vect{V}_{\rm i\pm1/2}^{n + 1/2} \approx \vect{V}_{\rm i\pm1/2}^{n} \,.
\end{equation}
However, this assumption is wrong even at first order in time, meaning that the error scales proportional to $\Delta t$. We implement this so-called Euler integrator in \ulula to demonstrate its instability (Fig.~\ref{fig:advection}).

To achieve a 2nd-order estimate, we evolve the reconstructed left and right states next to interfaces by a half time step. Of course, calculating the evolution is precisely the point of our solver, and we cannot know the eventual update to a cell's state before completing the Riemann solver calculation and differencing the fluxes across interfaces. However, we can get a first-order approximation of the evolution within a cell by differencing the fluxes that would arise from the reconstructed states at its left and right edges,
\begin{equation}
\label{eq:hancock_cons}
\vect{U}_{i \pm 1/2}^{n+1/2} = \vect{U}_{i \pm 1/2}^{n} + \frac{1}{2} \frac{\Delta t}{\Delta x} \left[ \mathcal{\fluxv}(\vect{V}_{i-1/2}^n) - \mathcal{\fluxv}(\vect{V}_{i+1/2}^n) \right] \,.
\end{equation}
We then convert back to primitive variables to obtain $\vect{V}_{i \pm 1/2}^{n+1/2}$, which serve as input to the Riemann solver at the respective interfaces. This update is known as the ``Hancock step.'' Note that if the slope $s_{\rm lim}$ was zero for a given variable, this update will not change the states because the edge states and fluxes will be the same. One disadvantage of this routine are the frequent primitive-conserved conversions: we reconstruct the edge states in $\vect{V}$, convert to $\vect{U}$, compute fluxes of to update $\vect{U}$, convert back to $\vect{V}$, and then apply the Riemann solver. \citet{mignone_10_ctuglm} note that the Hancock step can also be performed directly in primitive variables. Ignoring source terms and converting to primitive variables, we can write eq.~\ref{eq:euler_vector} in one dimension as 
\begin{equation}
\frac{\partial \vect{V}}{\partial t} = -\vect{A}_x \frac{\partial \vect{V}}{\partial x} \,,
\end{equation}
where $\vect{A}_x \equiv \dparinl{\fluxv (\vect{V})}{\vect{V}}$ is the Jacobian of the flux vector and the subscript $x$ reminds us that there are similar but different matrices for the other dimensions. However, since we are using a split solver where each sweep is executed in a single dimension, we can use the same matrix on vectors with interchanged velocity components (\S\ref{sec:implementation:timestepping}). For our 2D system of equations, the Jacobian turns out to be
\begin{equation}
\label{eq:jacobian}
\vect{A_x}(\vect{V}) =
\left(\begin{array}{cccc}
\vx & \rho & 0 & 0 \\ \noalign{\medskip}
0 & \vx & 0 & 1 / \rho \\ \noalign{\medskip}
0 & 0 & \vx & 0 \\ \noalign{\medskip}
0 & \cs^2 / \rho & 0 & \vx
\end{array}\right) \,.
\end{equation}
The Hancock step can now be performed in primitive variables as
\begin{equation}
\label{eq:hancock_prim}
\vect{V}_{i \pm 1/2}^{n+1/2} = \vect{V}_{i \pm 1/2}^{n} + \frac{\Delta t}{2}\ \vect{A}_x \left(\vect{V}_{i}^{n} \right)\ \vect{s}_{\rm lim} \,,
\end{equation}
since, by definition, $\vect{s}_{\rm lim}$ approximates $\dparinl{\vect{V}}{x}$. This algorithms gives virtually the same results as the conserved-variable Hancock step of eq.~\ref{eq:hancock_cons}, but it is faster and thus the default in \ulula. We could also use the reconstructed states at the cell edges to compute $\vect{A}_x(\vect{V}_{i\pm1/2}^{n})$, but this version makes virtually no difference in practice and demands two evaluations of the Jacobian.

A scheme that combines 2nd-order (linear) spatial interpolation and 2nd-order time interpolation via the Hancock step is often called a MUSCL-Hancock scheme, where the abbreviation stands for ``Monotonic Upstream-centered Scheme for Conservation Laws'' \citep{vanleer_79}. Our split version of this scheme has the desirable property that it is stable for any CFL number less than unity \citep{mignone_12_pluto}. We note that there are other, perhaps even simpler, ways to achieve second-order convergence in time. For example, in  Runge-Kutta integration schemes, multiple timesteps are performed and then averaged with certain weights. However, these schemes tend to be slower than MUSCL-Hancock due to the multiple evaluations of the Riemann fluxes.

\subsection{Gravity solver}
\label{sec:implementation:gravity}

\begin{figure*}
\centering
\includegraphics[trim =  18mm 20mm 3mm 8mm, clip, width=0.8\textwidth]{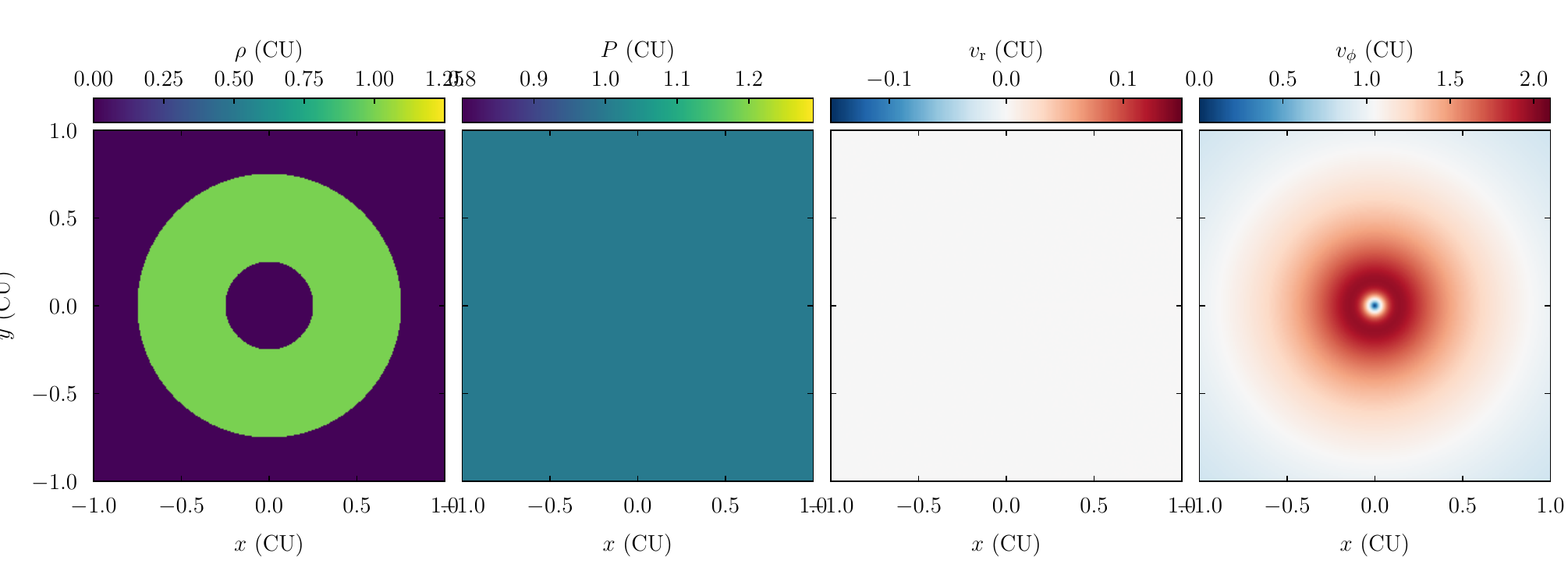}
\includegraphics[trim =  18mm 20mm 3mm 28mm, clip, width=0.8\textwidth]{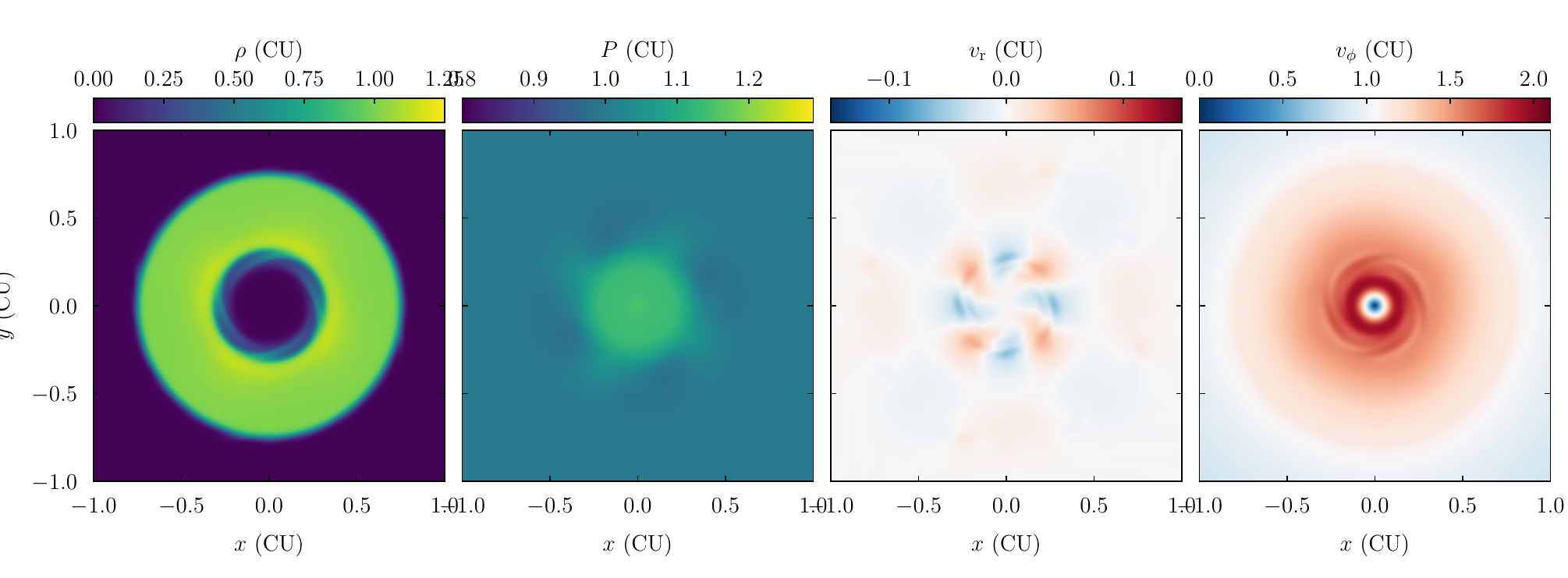}
\includegraphics[trim =  18mm 20mm 3mm 28mm, clip, width=0.8\textwidth]{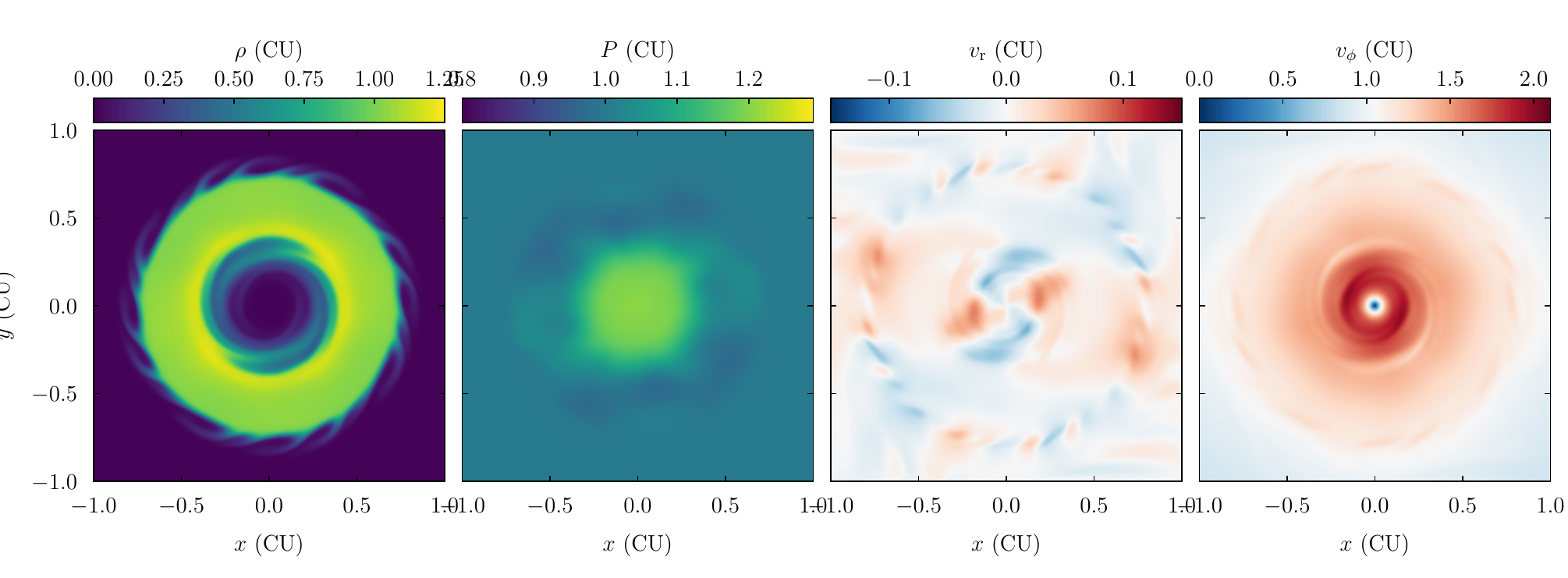}
\caption{Development of instabilities in a rotating Keplerian disk, shown at times $t = 0$, $5$, and $10$ in code units. The disk rotates under the fixed gravitational potential of a softened point mass that corresponds to the tangential velocity distribution shown in the top right panel. This setup should theoretically be static, but is a tough case for grid codes in practice. In our test, the disk becomes partially pressure-supported at the center, instead of gravity being purely balanced by its rotation.}
\label{fig:keplerian_disk}
\end{figure*}

\begin{figure*}
\centering
\includegraphics[trim =  18mm 20mm 2mm 8mm, clip, width=0.33\textwidth]{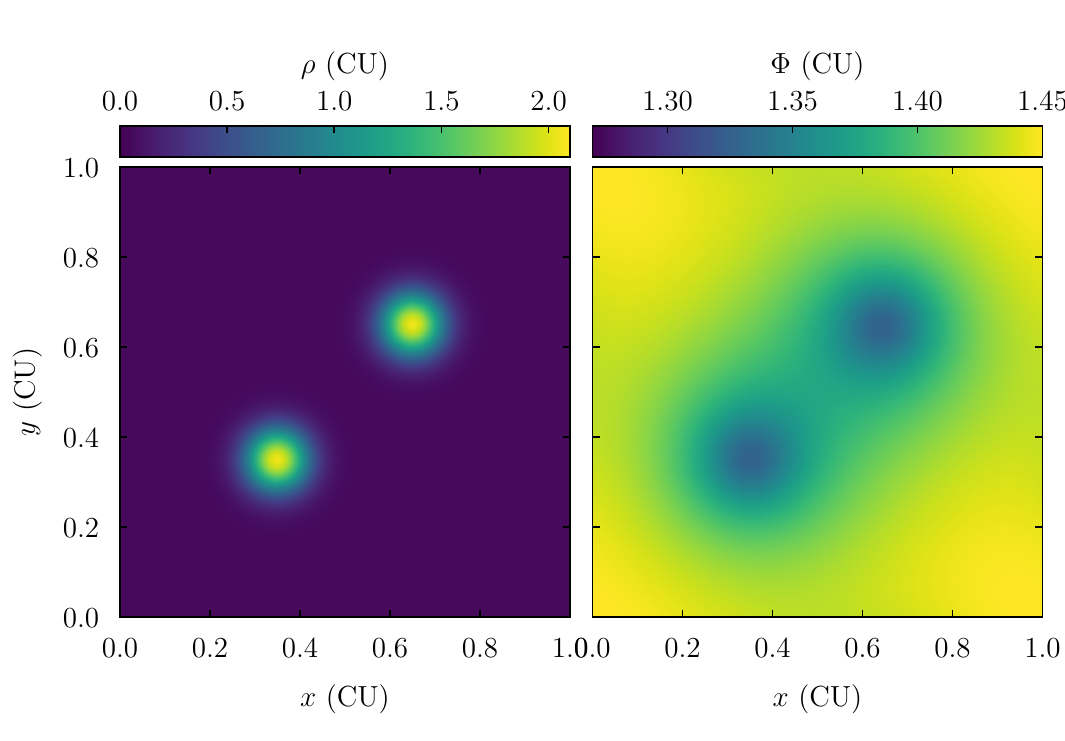}
\includegraphics[trim =  18mm 20mm 2mm 8mm, clip, width=0.33\textwidth]{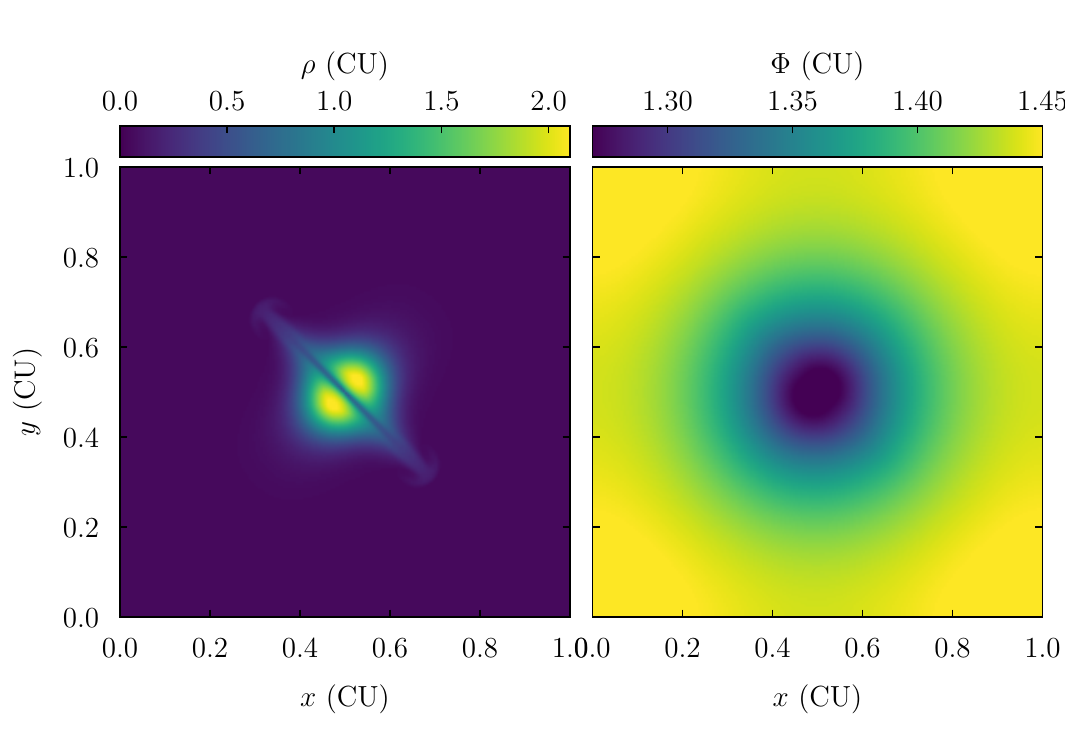}
\includegraphics[trim =  18mm 20mm 2mm 8mm, clip, width=0.33\textwidth]{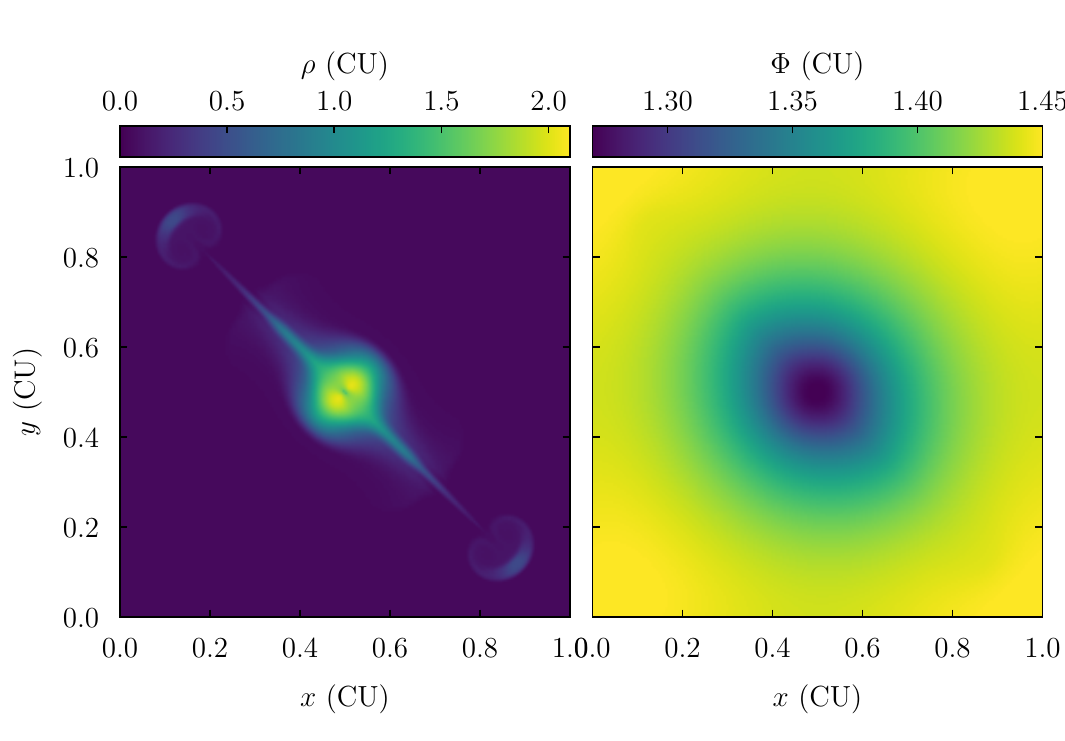}
\includegraphics[trim =  18mm 20mm 2mm 28mm, clip, width=0.33\textwidth]{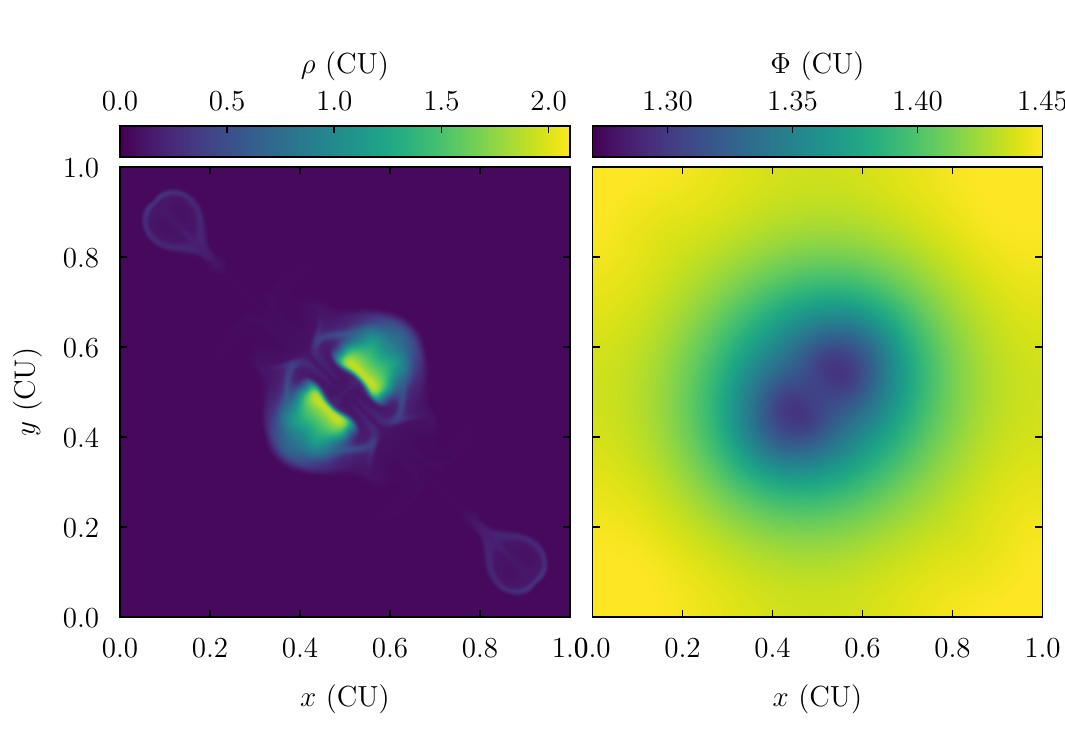}
\includegraphics[trim =  18mm 20mm 2mm 28mm, clip, width=0.33\textwidth]{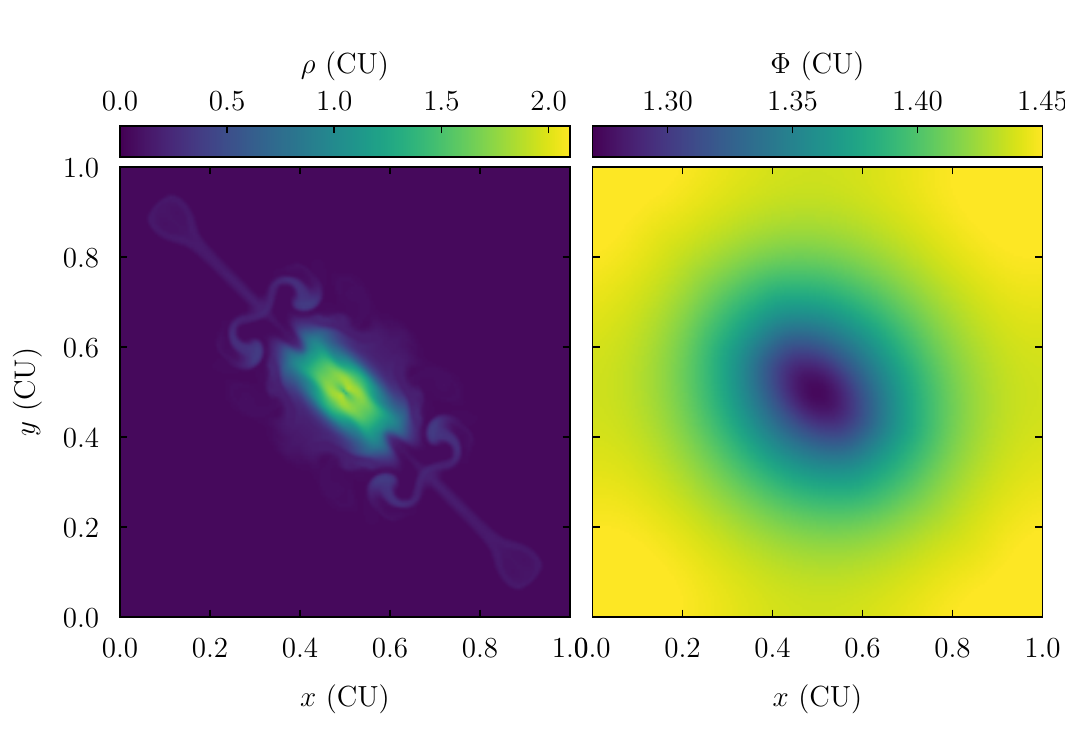}
\includegraphics[trim =  18mm 20mm 2mm 28mm, clip, width=0.33\textwidth]{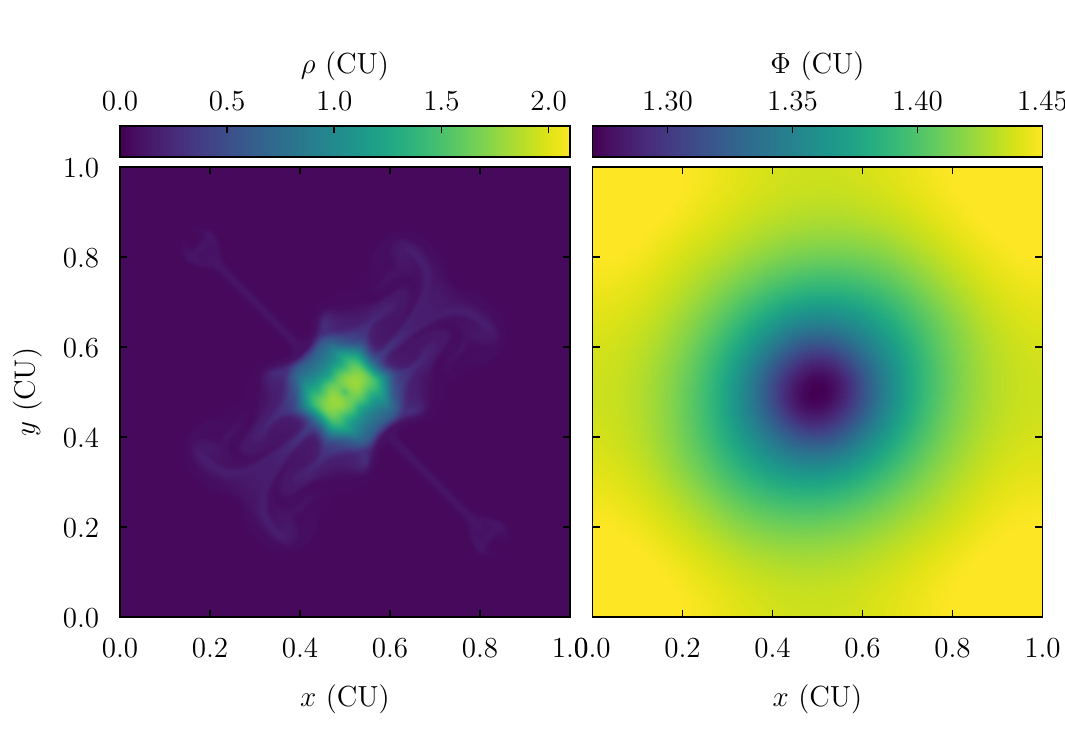}
\caption{Evolution of density and potential as two Gaussian gas blobs merge under their own self-gravity (in a periodic domain). The sets of panels show the simulation at times $t = 0$, $2$, $4$, $6$, $8$, and $10$. The interplay of pressure and gravity leads to complex, yet symmetric, structures.}
\label{fig:merger}
\end{figure*}

\ulula implements three kinds of gravitational potential. If a fixed acceleration $g$ is chosen, we interpret it as $\Phi = gy$ and add it as a source term in the $y$-momentum (or $x$ in the case of a 1D setup; see Fig.~\ref{fig:rt} for an example). If a constant, user-defined potential is set, we compute the gradients $\dparinl{\Phi}{x}$ and $\dparinl{\Phi}{y}$ once and subsequently apply them as a source term to the momenta (see Fig.~\ref{fig:keplerian_disk} for an example). In both cases, the source term for energy, $\dparinl{\Phi}{t}$, is zero. If self-gravity is activated, we compute the potential and its derivatives every time the source term is added (\S\ref{sec:implementation:timestepping}; see Fig.~\ref{fig:merger} for an example). Self-gravity works only with periodic boundary conditions because we solve the Poisson equation with a commonly used Fast Fourier Transform (FFT) technique. Specifically, we approximate the Laplacian via its 2nd-order discrete derivative,
\begin{align}
\label{eq:laplacian}
\lapl \Phi_{i,j} & \approx \frac{\Phi_{i+1,j} + \Phi_{i-1,j} + \Phi_{i,j+1} + \Phi_{i,j-1} - 4 \Phi_{i,j}}{\Delta x^2} \nonumber \\
& = 4 \pi G \rho_{ij} \,.
\end{align}
This equation can be solved exactly in Fourier space \citep[see e.g.][for a derivation]{zingale_21}. We multiply the FFT of the density field, $\widetilde{\rho}_{lm}$, by a Green's function to obtain the Fourier-space potential,
\begin{equation}
\label{eq:poisson_fft}
\widetilde{\Phi}_{lm} = \frac{2 \pi G\ \widetilde{\rho}_{lm}\  \Delta x^2}{\cos(2 \pi k_{x,lm}) + \cos(2 \pi k_{y,lm}) - 2} \,,
\end{equation}
to which we apply an inverse FFT to obtain $\Phi_{ij}$. The indices $l$ and $m$ now run over the vectors $k_x$ and $k_y$, which here refer to the dimensionless wavenumbers used by the FFT routines,
\begin{equation}
k_x = \left[0, \frac{1}{N_x}, \frac{2}{N_x}, ..., 0.5-\frac{1}{N_x}, -0.5, -0.5 + \frac{1}{N_x}, ..., -\frac{1}{N_x} \right] \,, \nonumber
\end{equation}
where $N_x$ is the number of cells in the $x$-direction, and similarly for $k_y$. The Green's function for the $l=m=0$ term is undefined. This term is known as the ``DC mode'' and corresponds to a uniform background density \citep[e.g.,][]{gnedin_11_dc}. While the normalization of the input density directly translates into the normalization of $\Phi$ and thus $\lapl \Phi$, a uniform offset in a periodic domain corresponds to an infinite extent of mass that has no net gravitational effect. Since there is no well-defined zero point for the potential, it is not clear how to translate the DC mode into a potential. For this term only, we set the denominator of Equation~\ref{eq:poisson_fft} to unity. We have checked that the Laplacian as defined in Equation~\ref{eq:laplacian} matches the right-hand side of the Poisson equation to high accuracy (after subtracting the mean, or DC mode, from the latter). \ulula also allows the user to combine self-gravity with an additive, fixed potential.

\subsection{Code structure and features}
\label{sec:implementation:code}

Besides the actual hydro solver, \ulula has a number of components that make it easy to set up, run, and plot simulations.  
\begin{itemize}
\setlength{\itemsep}{1pt}

\item {\bf Object-oriented structure:} The core of \ulula consists of four units. The solver algorithms described in \S\ref{sec:implementation} are contained in a single class called \texttt{Simulation}. Physical problem setups are classes derived from a base class called \texttt{Setup} (details to follow in \S\ref{sec:results:setups}). A plotting and analysis module extracts quantities from the simulation. Finally, a ``driver'' routine called \texttt{run} runs a Setup, given parameters such as the final time or a maximum number of timesteps, the hydro solver to be used, and desired outputs. Running a simulation can thus be achieved with as few as two lines of python code.

\item {\bf Plotting and movies:} A separate plotting module produces one-dimensional or two-dimensional (colormap) plots of a given set of fluid properties. When output times do not exactly match a timestep, the simulation is run to the exact desired output time. \ulula can also produce movies of a given duration, resolution, and frame rate, either in the \texttt{mp4} format (using the \texttt{ffmpeg} library) or as a  \texttt{gif}.

\item {\bf Units:} Given that the Euler equations are scale-free, the \texttt{Simulation} module works in an abstracted system of code units that can represent any length, time, and mass scale (or combinations thereof such as density). The user can freely set those three scales, but they are never used in the actual simulation module. They are, however, used when plotting, since the user can choose between a wide range of length, time, and mass units to plot (which do not need to match the code units of the simulation). 

\item {\bf Boundary conditions:} \ulula implements periodic, outflow, and wall boundary conditions. The user can also set callback functions that allow for setting user-defined boundary conditions, or even interacting with the fluid state at runtime. 

\item {\bf File I/O:} An \ulula simulation can be saved to and loaded from \texttt{hdf5} files.
\end{itemize}

The total code volume of \ulula is just under $3000$ active lines of python code, about $1200$ of which are dedicated to specific problem setups.


\section{Problem setups and code tests}
\label{sec:results}

We begin by briefly describing the pre-implemented problem setups in \S\ref{sec:results:setups}. We then use some of those setups to assess the numerical convergence of \ulula in \S\ref{sec:results:accuracy}. We measure the code's performance in \S\ref{sec:results:performance}.

\subsection{Pre-implemented problem setups}
\label{sec:results:setups}

\begin{figure}
\centering
\includegraphics[trim =  2mm 3mm 3mm 0mm, clip, width=0.38\textwidth]{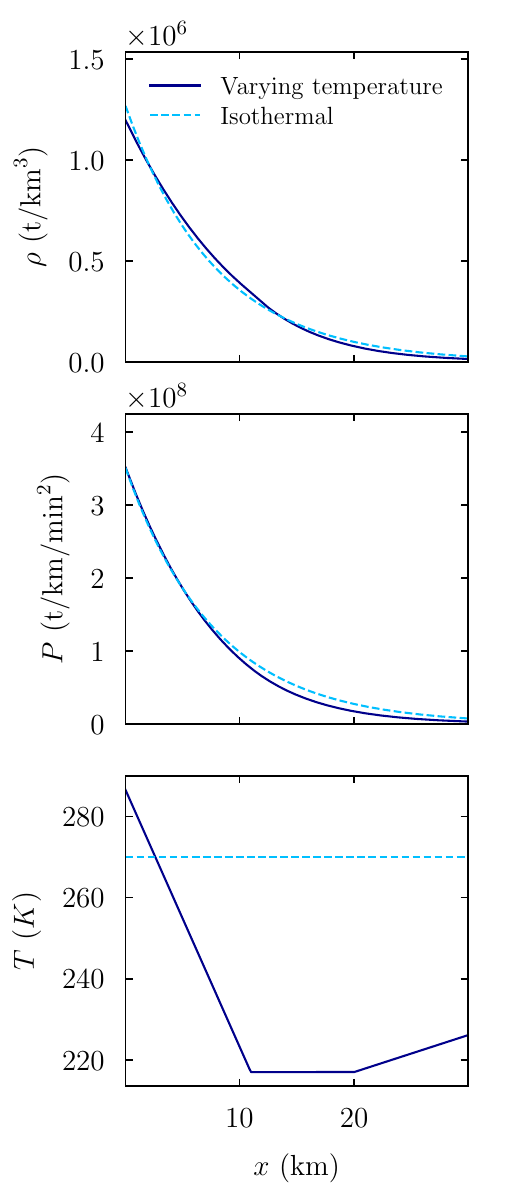}
\caption{Vertical density, pressure, and temperature profiles of a hydrostatic atmosphere for the conditions on Earth. The atmosphere is modeled using wall boundary conditions at $0$ and $30\ \km$ altitude. The problem can be solved analytically for an isothermal equation of state, resulting in an exponential density and pressure profile (dashed light-blue lines). The dark blue line in the bottom panel represents an approximation to the true temperature variations in Earth's atmosphere. The resulting deviations in density and pressure are relatively small, validating the commonly used isothermal approximation. The figure also demonstrates the arbitrary unit system in which \ulula results can be plotted, which was set to kilometers, tons, and minutes in this case.}
\label{fig:atmosphere}
\end{figure}

As described in \S\ref{sec:implementation:code}, all code to set up a particular hydrodynamical problem is bundled in class objects in \ulula, which are derived from a common base class. In the simplest case, the user may overwrite only two functions that give a brief string name for the setup and set the initial fluid state, (possibly including an EOS, boundary conditions, and so on). The setup class also contains more advanced routines that can be used to provide a known solution to plotting routines, to set ranges for plots, or to interact with the simulation at runtime. While creating new setups is an important part of the learning process, a number of setups are pre-implemented in \ulula. They include well-known tests of hydro codes and algorithms as well as commonly discussed physics such as sound waves and instabilities. Specifically, the following one-dimensional problems are implemented.
\begin{itemize}
\setlength{\itemsep}{1pt}

\item {\bf 1D Advection:} a tophat or sine wave density distribution is moved through a 1D domain at fixed velocity. This possibly simplest test of any hydro solver reveals a lot about the impact of interpolation and slope limiters.

\item {\bf Sound wave:} a soundwave with user-defined frequency and amplitude is imposed as a boundary condition on the left edge of the domain. The wave travels in an ideal or isothermal gas, demonstrating the existence of sound waves, their speed, and wave steepening.

\item {\bf Shock tube:} the famous \citet{sod_78} shock test primarily tests the accuracy of Riemann solvers.

\item {\bf Free-fall:} a Gaussian density distribution falls under fixed-acceleration gravity in an open domain. This setup is designed to test the accuracy of the gravitational source terms.

\item {\bf Atmosphere:} a classic problem in hydrostatics is the structure of an isothermal atmosphere. The \ulula setup corresponds to Earth's atmosphere, which takes a surprisingly long time to settle into the expected exponential profile in density and pressure (partly due to the wall boundary conditions). The \ulula setup has an option to use a more realistic temperature profile to explore the impact of the isothermal assumption (Fig.~\ref{fig:atmosphere}).

\item {\bf Jeans instability:} a sinusoidal density perturbation with user-defined frequency and amplitude is evolved under self-gravity. The user chooses a Jeans length that determines whether the perturbation collapses or is supported by pressure.

\end{itemize}

The following two-dimensional problems are implemented:

\begin{itemize}
\setlength{\itemsep}{1pt}

\item {\bf 2D Advection:} similar to the 1D advection setup, but advecting a 2D tophat or sine wave pattern (Fig.~\ref{fig:advection}).

\item {\bf Kelvin-Helmholtz instability:} the well-known instability at the interface of two fluids flowing past each other. The user can choose between sharp or smooth initial conditions \citep[e.g.,][]{robertson_10} and the wavelength of the initial velocity perturbation.

\item {\bf Cloud crushing:} a cold cloud is disrupted by a hot, fast wind \citep[e.g.,][]{klein_94}. This type of setup can give very different results that depend on the hydro algorithm \citep{agertz_07} and the physics included \citep[e.g.,][]{gronke_18}.

\item {\bf Sedov-Taylor explosion:} in this classic blast wave setup \citep{taylor_50, sedov_59}, a large amount of energy is injected at the center of the domain. The energy is distributed over a narrow Gaussian kernel to avoid resolution effects. \ulula matches the known analytical solution.

\item {\bf Gresho vortex:} a radial pressure profile is balanced by rotation, providing a stringent test for angular momentum conservation that grid codes often struggle with \citep{gresho_87, liska_03}. \ulula performs well on this test (\S\ref{sec:results:accuracy}).

\item {\bf Rayleigh-Taylor instability:} a well-known fluid instability where a heavier fluid initially rests on top of a lighter fluid (Fig.~\ref{fig:rt}). Linear theory makes predictions for the initial evolution of the density perturbation \citep[e.g.,][]{clarke_14}, which can be verified with \ulula.

\item {\bf Tidal disruption:} a Gaussian gas blob orbits in the potential of a point mass and is disrupted.

\item {\bf Keplerian disk:} a gas disk rotates in the gravitational potential of a softened point mass, $\Phi = - (r^2 + \epsilon^2)^{-1/2}$. The rotational velocity is set such that it balances the potential, $v_\rmr = r (r^2 + \epsilon^2)^{-3/4}$. This setup should theoretically be stable and keep rotating forever, but in practice, grid codes struggle to preserve the angular momentum and sharp edges of the disk (Fig.~\ref{fig:keplerian_disk}).

\item {\bf Merger:} two Gaussian gas blobs collide under their own gravity. Following their collision, the interplay of pressure and gravity creates complex, symmetric patterns that eventually settle down into a single blob of gas (Fig.~\ref{fig:merger}).

\end{itemize}

While many of these setups come with certain user-adjustable parameters, there are merely examples of possible implementations, and they are intended to be modified and extended. For example, what would happen if we inject a sound wave that is not sinusoidal? How does tidal disruption depend on the shape of the gravitational potential? How does the Sedov-Taylor explosion proceed if the background density is not uniform? All of these questions could be answered by adding a few extra lines of code.

\subsection{Convergence}
\label{sec:results:accuracy}

\begin{figure}
\centering
\includegraphics[trim =  4mm 8mm 5mm 0mm, clip, width=0.4\textwidth]{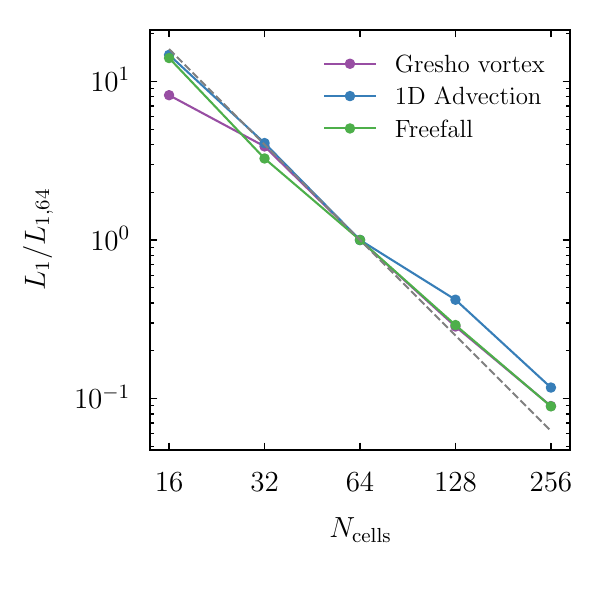}
\caption{Convergence with resolution in the Gresho Vortex, 1D sine wave advection, and free-fall tests. We compare the density solution after one code time unit to the equilibrium (initial) solution and take the absolute fractional error ($L_1$ norm), normalized at $N_{\rm cells} = 64$ to make the tests more easily comparable. The gray line shows a slope of $-2$ as would be expected for a perfect second-order solver. \ulula almost, though not quite, achieves second-order convergence, including in the challenging Gresho vortex test case and when gravity is present.}
\label{fig:convergence}
\end{figure}

\ulula's runtime function automatically checks the conservation of mass, momentum, and energy, as long as those quantities are expected to be conserved. Exceptions include outflow boundary conditions (which violate all conservation laws) and wall boundary conditions (which violate only momentum conservation). The conservation laws are generally obeyed to machine precision in setups without gravity (since the gravitational source terms in Equation~\ref{eq:vectors} violate the strict conservation laws).

A number of the setups described in \S\ref{sec:results:setups} have known solutions, which can be plotted by \ulula for comparison. For example, the 1D and 2D advection setups test the diffusivity of different schemes (Fig.~\ref{fig:advection}). The shock-tube and Sedov-Taylor setups test the shock-capturing properties of the scheme, especially of the Riemann solver. The Gresho vortex and Keplerian disk setups test angular momentum conservation and are challenging tests for grid codes. 

In Fig.~\ref{fig:convergence} we show the numerical convergence with resolution in the 1D advection, free-fall, and Gresho vortex setups. In all cases, we find convergence rates that are slightly less steep than $N^{-2}$ (gray line), which would be expected for a solver with perfect 2nd-order convergence. Nonetheless, the tests demonstrate that the MUSCL-Hancock scheme laid out in \S\ref{sec:implementation} basically converges as expected, including in the presence of gravitational source terms.

\subsection{Performance}
\label{sec:results:performance}

\begin{figure}
\centering
\includegraphics[trim =  0mm 0mm 0mm 0mm, clip, width=0.47\textwidth]{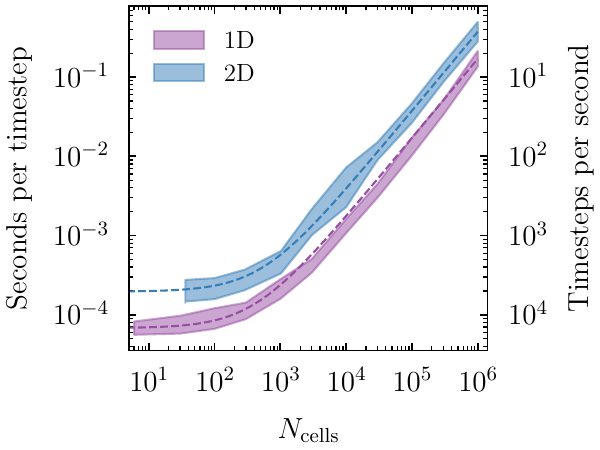}
\caption{Performance of \ulula on a standard laptop, expressed as the CPU time taken per timestep. The $x$-axis shows the problem size as the total number of cells (including ghost cells). The shaded areas are delineated by the times for the computationally cheapest and most expensive types of problems and hydro solvers, namely a first-order Euler scheme with HLL and the full MUSCL-Hancock scheme with HLLC and gravity. The time taken is well approximated by an overhead per timestep (about $2 \times 10^{-4}\ \scnd$ in 2D) plus a fixed time per cell, which is what we would expect for perfect scaling with problem size (about $0.4$ seconds per million cells in 2D). Both the overhead and time per timestep are larger for 2D simulations, which is expected due to the multiple dimensional sweeps.}
\label{fig:performance}
\end{figure}

One possible concern about python as a programming language could be that it is notoriously slow, which could lead to computing power being dedicated to the overhead of the code rather than solving the Euler equations. All array operations in \ulula use the \texttt{numpy} library and are thus implicitly implemented in \texttt{C}. Fig.~\ref{fig:performance} shows the code's performance as measured on a 2025 MacBook Pro. We test 1D and 2D setups, both for a minimal calculation (isothermal, HLL Riemann solver, and so on) and a demanding calculation (gravity, HLLC, and so on). The shaded regions lie between those cases and should thus encompass most realistic scenarios. 

The run time is well approximated by a constant overhead per timestep plus a fixed time per cell, corresponding to the operations of the python framework and the \texttt{numpy} array operations, respectively. Reassuringly, the overhead is negligible for all but the smallest problems, about $2 \times 10^{-4}\ \scnd$ per timestep in 2D, corresponding to $5000$ timesteps per second. For any sizable problem, \ulula's runtime thus scales linearly as about $4 \times 10^{-7}$ seconds per cell. In 1D, this time is roughly halved to $2 \times 10^{-7}$ seconds per cell, which makes sense given that we execute one instead of two dimensional sweeps. We conclude that \ulula is limited by the speed of the \texttt{numpy} library. 

\section{Conclusions and future development}
\label{sec:future}

We have presented \ulula, an ultra-lightweight 2D python hydro solver for teaching and experimentation. The main conclusion is that it is possible to create such a framework in less than $800$ active lines of active code without compromising on the fidelity of the hydrodynamical algorithms or the execution speed, which is essentially limited by the execution time of \texttt{numpy} operations on a single CPU. 

\ulula is not intended as a monolithic hydro solver but as a platform for quick experimentation with a low barrier to entry. There are numerous ways in which the code could be improved and expanded. Some examples include:

\begin{itemize}
\setlength{\itemsep}{1pt}

\item {\bf Elements of the solver:} There are many more options for the various elements of \ulula's HD scheme, for example the \citet{roe_81} Riemann solver, additional slope limiters (\S\ref{sec:implementation:spatial}), or other time integration schemes such as Runge-Kutta. 

\item {\bf Higher-order schemes:} The algorithms in \ulula were chosen to achieve 2nd-order convergence. Given that it is easy to increase the number of ghost cells in \ulula, there are no barriers to implementing higher-order algorithms such at the piecewise-parabolic method \citep{colella_84}.

\item {\bf Geometry:} \ulula knows only 1D and 2D cartesian grids, but many problems are better solved in spherically symmetric or 2D cylindrical coordinates. Such geometries would add certain terms to the equations.

\item {\bf Viscosity:} One of the most important simplifications inherent in the Euler equations as written in eq.~\ref{eq:euler} is to neglect viscosity. One could add simplified formulations, e.g., isotropic bulk and shear velocity, as additional source terms.

\item {\bf Cooling:} In astrophysics, many important applications depend on radiative cooling, which could be implemented as an additional source term. The timestep would need to be limited such that no more than some fraction of the internal energy can be subtracted during a timestep.

\item {\bf Parallelization:} Currently, \ulula runs on a single CPU. A true, multi-node parallelization method such as MPI would lead to a significant increase in code complexity (e.g., due to domain decomposition). However, multi-threading would be a more lightweight avenue towards accelerating the code, e.g., using libraries such as \texttt{jax} or \texttt{numba}. 

\end{itemize}

This list deliberately avoids physics that behaves intrinsically differently in 3D. Most notably, one could imagine extending \ulula to MHD, for example using divergence-cleaning schemes that work only with cell-centered quantities \citep{dedner_02, mignone_10_ctuglm}. However, the majority of interesting MHD problems are intrinsically 3D. Similarly, turbulence behaves fundamentally differently in 3D, which is why we have not added a turbulent setup. None the less, the list of suggestions above is highly incomplete, and we hope that \ulula will inspire all kinds of numerical experimentation. 


\section*{Acknowledgments}

I thank Philip Mocz for insightful comments on a draft. I am grateful for certain code contributions to \ulula, namely to Calvin Osinga (for the routine to create gif videos), to Philip Mansfield (for the annulus-square overlap routine used in radial plotting), and for the exact solution to the Sedov-Taylor explosion problem, which was adapted from the Gandalf code by \citet{hubber_18}. The exact solution to the shock tube problem was based on the excellent lecture notes by Frank van den Bosch and Susanne Höfner. \ulula was developed while teaching ASTR 670, graduate class in hydrodynamics at the University of Maryland. Many thanks to all the students who bore with my experimentation with \ulula in class, and who experimented with the code themselves! This research was supported in part by the National Science Foundation under Grant number 2338388. I am furthermore grateful to the Sloan Foundation for their financial support. \ulula makes extensive use of the python packages \textsc{Numpy} \citep{code_numpy2}, \textsc{Scipy} \citep{code_scipy}, and \textsc{Matplotlib} \citep{code_matplotlib}.

\vspace{0.5cm}




\bibliographystyle{\includedir/citestyle_mnras2}
\bibliography{\includedir/bib_mine.bib,\includedir/bib_general.bib,\includedir/bib_structure.bib,\includedir/bib_galaxies.bib,\includedir/bib_clusters.bib,\includedir/bib_hydro_mhd_cr.bib}

\begin{thebibliography}{40}
\expandafter\ifx\csname natexlab\endcsname\relax\def\natexlab#1{#1}\fi

\bibitem[{{Agertz} {et~al}\mbox{.}(2007){Agertz}, {Moore}, {Stadel}, {Potter},
  {Miniati}, {Read}, {Mayer}, {Gawryszczak}, {Kravtsov}, {Nordlund}, {Pearce},
  {Quilis}, {Rudd}, {Springel}, {Stone}, {Tasker}, {Teyssier}, {Wadsley}, \&
  {Walder}}]{agertz_07}
{Agertz} O. {et~al.}, 2007, \mnras, 380, 963

\bibitem[{{Batten} {et~al}\mbox{.}(1997){Batten}, {Clarke}, {Lambert}, \&
  {Causon}}]{batten_97}
{Batten} P., {Clarke} N., {Lambert} C., {Causon} D.~M., 1997, SIAM Journal on
  Scientific Computing, 18, 1553

\bibitem[{{Burns} {et~al}\mbox{.}(2020){Burns}, {Vasil}, {Oishi}, {Lecoanet},
  \& {Brown}}]{burns_20}
{Burns} K.~J., {Vasil} G.~M., {Oishi} J.~S., {Lecoanet} D., {Brown} B.~P.,
  2020, Physical Review Research, 2, 023068

\bibitem[{{Clarke} \& {Carswell}(2014)}]{clarke_14}
{Clarke} C., {Carswell} B., 2014, {Principles of Astrophysical Fluid Dynamics}.
  Cambridge University Press

\bibitem[{{Colella} \& {Glaz}(1985)}]{colella_85_cg}
{Colella} P., {Glaz} H.~M., 1985, Journal of Computational Physics, 59, 264

\bibitem[{{Colella} \& {Woodward}(1984)}]{colella_84}
{Colella} P., {Woodward} P.~R., 1984, Journal of Computational Physics, 54, 174

\bibitem[{{Courant}, {Friedrichs} \& {Lewy}(1928){Courant}, {Friedrichs}, \&
  {Lewy}}]{courant_28}
{Courant} R., {Friedrichs} K., {Lewy} H., 1928, Mathematische Annalen, 100, 32

\bibitem[{{Dedner} {et~al}\mbox{.}(2002){Dedner}, {Kemm}, {Kr{\"o}ner}, {Munz},
  {Schnitzer}, \& {Wesenberg}}]{dedner_02}
{Dedner} A., {Kemm} F., {Kr{\"o}ner} D., {Munz} C.~D., {Schnitzer} T.,
  {Wesenberg} M., 2002, Journal of Computational Physics, 175, 645

\bibitem[{{Gnedin}, {Kravtsov} \& {Rudd}(2011){Gnedin}, {Kravtsov}, \&
  {Rudd}}]{gnedin_11_dc}
{Gnedin} N.~Y., {Kravtsov} A.~V., {Rudd} D.~H., 2011, \apjs, 194, 46

\bibitem[{Godunov(1959)}]{godunov_59}
Godunov S.~K., 1959, Matematicheskii Sbornik, 47, 271, english translation in:
  J. Comput. Phys., vol. 3, 1965, pp. 251--266

\bibitem[{Gresho \& Sani(1987)}]{gresho_87}
Gresho P.~M., Sani R.~L., 1987, Incompressible Flow and the Finite Element
  Method. Wiley, New York

\bibitem[{{Gronke} \& {Oh}(2018)}]{gronke_18}
{Gronke} M., {Oh} S.~P., 2018, \mnras, 480, L111

\bibitem[{{Harpole} {et~al}\mbox{.}(2019){Harpole}, {Zingale}, {Hawke}, \&
  {Chegini}}]{harpole_19}
{Harpole} A., {Zingale} M., {Hawke} I., {Chegini} T., 2019, The Journal of Open
  Source Software, 4, 1265

\bibitem[{{Harris} {et~al}\mbox{.}(2020){Harris}, {Millman}, {van der Walt},
  {Gommers}, {Virtanen}, {Cournapeau}, {Wieser}, {Taylor}, {Berg}, {Smith},
  {Kern}, {Picus}, {Hoyer}, {van Kerkwijk}, {Brett}, {Haldane}, {del R{\'\i}o},
  {Wiebe}, {Peterson}, {G{\'e}rard-Marchant}, {Sheppard}, {Reddy}, {Weckesser},
  {Abbasi}, {Gohlke}, \& {Oliphant}}]{code_numpy2}
{Harris} C.~R. {et~al.}, 2020, \nat, 585, 357

\bibitem[{Harten, Lax \& van Leer(1983)Harten, Lax, \& van Leer}]{harten_83}
Harten A., Lax P.~D., van Leer B., 1983, SIAM Review, 25, 35

\bibitem[{{Hubber}, {Rosotti} \& {Booth}(2018){Hubber}, {Rosotti}, \&
  {Booth}}]{hubber_18}
{Hubber} D.~A., {Rosotti} G.~P., {Booth} R.~A., 2018, \mnras, 473, 1603

\bibitem[{Hunter(2007)}]{code_matplotlib}
Hunter J.~D., 2007, Computing in Science Engineering, 9, 90

\bibitem[{{Klein}, {McKee} \& {Colella}(1994){Klein}, {McKee}, \&
  {Colella}}]{klein_94}
{Klein} R.~I., {McKee} C.~F., {Colella} P., 1994, \apj, 420, 213

\bibitem[{{Li}(2005)}]{li_05_hllc}
{Li} S., 2005, Journal of Computational Physics, 203, 344

\bibitem[{{Lien} \& {Leschziner}(1994)}]{lien_94}
{Lien} F.~S., {Leschziner} M.~A., 1994, International Journal for Numerical
  Methods in Fluids, 19, 527

\bibitem[{{Liska} \& {Wendroff}(2003)}]{liska_03}
{Liska} R., {Wendroff} B., 2003, SIAM Journal on Scientific Computing, 25, 995

\bibitem[{{Mignone} \& {Tzeferacos}(2010)}]{mignone_10_ctuglm}
{Mignone} A., {Tzeferacos} P., 2010, Journal of Computational Physics, 229,
  2117

\bibitem[{{Mignone} {et~al}\mbox{.}(2012){Mignone}, {Zanni}, {Tzeferacos}, {van
  Straalen}, {Colella}, \& {Bodo}}]{mignone_12_pluto}
{Mignone} A., {Zanni} C., {Tzeferacos} P., {van Straalen} B., {Colella} P.,
  {Bodo} G., 2012, The Astrophysical Journal Supplement Series, 198, 7

\bibitem[{{Ramachandran} {et~al}\mbox{.}(2019){Ramachandran}, {Bhosale},
  {Puri}, {Negi}, {Muta}, {Dinesh}, {Menon}, {Govind}, {Sanka}, {Sebastian},
  {Sen}, {Kaushik}, {Kumar}, {Kurapati}, {Patil}, {Tavker}, {Pandey},
  {Kaushik}, {Dutt}, \& {Agarwal}}]{ramachandran_19}
{Ramachandran} P. {et~al.}, 2019, arXiv e-prints, arXiv:1909.04504

\bibitem[{{Robertson} {et~al}\mbox{.}(2010){Robertson}, {Kravtsov}, {Gnedin},
  {Abel}, \& {Rudd}}]{robertson_10}
{Robertson} B.~E., {Kravtsov} A.~V., {Gnedin} N.~Y., {Abel} T., {Rudd} D.~H.,
  2010, \mnras, 401, 2463

\bibitem[{Roe(1981)}]{roe_81}
Roe P.~L., 1981, Journal of Computational Physics, 43, 357

\bibitem[{{Roe}(1986)}]{roe_86}
{Roe} P.~L., 1986, Annual Review of Fluid Mechanics, 18, 337

\bibitem[{{Sedov}(1959)}]{sedov_59}
{Sedov} L.~I., 1959, {Similarity and Dimensional Methods in Mechanics}. CRC
  Press

\bibitem[{{Sod}(1978)}]{sod_78}
{Sod} G.~A., 1978, Journal of Computational Physics, 27, 1

\bibitem[{{Strang}(1968)}]{strang_68}
{Strang} G., 1968, SIAM Journal on Numerical Analysis, 5, 506

\bibitem[{{Taylor}(1950)}]{taylor_50}
{Taylor} G., 1950, Proceedings of the Royal Society of London Series A, 201,
  159

\bibitem[{{Toro}(2009)}]{toro_09}
{Toro} E., 2009, {Riemann solvers and numerical methods for fluid dynamics},
  3rd edn. Springer

\bibitem[{{Toro}, {Spruce} \& {Speares}(1994){Toro}, {Spruce}, \&
  {Speares}}]{toro_94}
{Toro} E.~F., {Spruce} M., {Speares} W., 1994, Shock Waves, 4, 25

\bibitem[{{van Albada}, {van Leer} \& {Roberts}(1982){van Albada}, {van Leer},
  \& {Roberts}}]{vanalbada_82}
{van Albada} G.~D., {van Leer} B., {Roberts}, W.~W. J., 1982, \aap, 108, 76

\bibitem[{{van Leer}(1974)}]{vanleer_74}
{van Leer} B., 1974, Journal of Computational Physics, 14, 361

\bibitem[{{van Leer}(1977)}]{vanleer_77}
{van Leer} B., 1977, Journal of Computational Physics, 23, 263

\bibitem[{{van Leer}(1979)}]{vanleer_79}
{van Leer} B., 1979, Journal of Computational Physics, 32, 101

\bibitem[{{Virtanen} {et~al}\mbox{.}(2020){Virtanen}, {Gommers}, {Oliphant},
  {Haberland}, {Reddy}, {Cournapeau}, {Burovski}, {Peterson}, {Weckesser},
  {Bright}, {van der Walt}, {Brett}, {Wilson}, {Millman}, {Mayorov}, {Nelson},
  {Jones}, {Kern}, {Larson}, {Carey}, {Polat}, {Feng}, {Moore}, {VanderPlas},
  {Laxalde}, {Perktold}, {Cimrman}, {Henriksen}, {Quintero}, {Harris},
  {Archibald}, {Ribeiro}, {Pedregosa}, {van Mulbregt}, \& {SciPy 1. 0
  Contributors}}]{code_scipy}
{Virtanen} P. {et~al.}, 2020, Nature Methods, 17, 261

\bibitem[{{Zingale}(2014)}]{zingale_14_pyro}
{Zingale} M., 2014, Astronomy and Computing, 6, 52

\bibitem[{{Zingale}(2021)}]{zingale_21}
{Zingale} M., 2021, { Introduction to Computational Astrophysical
  Hydrodynamics}. The Open Astrophysics Bookshelf

\end{thebibliography}




\ifuseapj
\else
\bsp
\label{lastpage}
\fi

\end{document}